\documentclass[sn-mathphys,Numbered, numbers]{sn-jnl}% Math and Physical Sciences Reference Style
\usepackage{graphicx}%
\usepackage{multirow}%
\usepackage{amsmath,amssymb,amsfonts}%
\usepackage{amsthm}%
\usepackage{mathrsfs}%
\usepackage[title]{appendix}%
\usepackage{xcolor}%
\usepackage{textcomp}%
\usepackage{manyfoot}%
\usepackage{booktabs}%
\usepackage{algorithmicx}%
\usepackage{algpseudocode}%
\usepackage{listings}%
\usepackage{float}
\usepackage{subfig}
\theoremstyle{thmstyleone}%
\theoremstyle{thmstyletwo}%

\theoremstyle{thmstylethree}%

\usepackage[linesnumbered,ruled,vlined]{algorithm2e}
\DeclareGraphicsExtensions{.png,.pdf}
\usepackage{enumitem}

\usepackage{url}

\algnewcommand\Break{\textbf{break}}
\graphicspath{{./figure/}}

\begin{document}

\title[Transparent Invalidation]{The Case of Transparent Cache Invalidation in Web Applications }

\author[1]{\fnm{Yunhong} \sur{ Ji}}\email{jiyunhong@ruc.edu.cn}

\author*[2]{\fnm{Xuan} \sur{Zhou}}\email{xzhou@dase.ecnu.edu.cn}

\author[3]{\fnm{Yongluan} \sur{Zhou}}\email{zhou@di.ku.dk}

\author[2]{\fnm{Ke} \sur{Wang}}\email{52275903009@stu.ecnu.edu.cn}

\affil*[1]{ \orgdiv{School of Information}, \orgname{Renmin University of China}, \orgaddress{\city{Beijing},  \postcode{100872}, \country{China}}}

\affil[2]{\orgdiv{School of Data Science and Engineering},\orgname{East China Normal University}, \orgaddress{\city{Shanghai}, \postcode{200062}, \country{China}}}

\affil[3]{\orgdiv{Department of Computer Science}, \orgname{University of Copenhagen}, \orgaddress{\city{Copenhagen}, \postcode{DK-2100}, \country{Denmark}}}

%\renewcommand{\shortauthors}{Ji and Zhou, et al.}

\abstract{ Application-level caches are widely adopted by web applications to minimize the response time of user requests as well as to reduce the burden on the system backend, such as the database servers. In the state of practice, developers have to take care of the data freshness of application-level caches manually. Given the growing complexities of today's web applications, it becomes increasingly challenging for developers to understand, reason about, and implement cache invalidation methods.  Furthermore, according to our survey of open-source web application projects and engineers, it is indeed challenging to map database updates with cache entries at the application level. Therefore, we propose a design to handle data validity in a transparent and precise manner, without requiring any intervention from developers. Its main idea is to modify the DBMS to provide necessary information for cache management and enhance the cache with an invalidation index to identify and invalidate outdated data automatically and efficiently. Based on the design, we further provide two specific solutions. Our preliminary experiments indicate that our solutions could effectively achieve transparent cache invalidation while maintaining cost-effectiveness.}

\keywords{Application-level cache, cache invalidation, interval tree, bloom filter}

\maketitle

\section{Introduction}

Application-level caches are crucial for improving the performance of modern web applications~\cite{Survey_Application, Study_Application}. Typically, by storing results of frequently invoked methods operating over databases, they effectively reduce response latency, elevate user experience, and alleviate backend system workload~~\cite{VLDB_TUTORIAL, ChronoCache}.  Ideally, the management of the application-level cache should be transparent, allowing developers to concentrate on core business logic without explicit handling of cache contents. However, the current landscape of application-level cache management remains intricate, necessitating developers to possess in-depth and up-to-date knowledge of cache contents~\cite{Survey_Application}. 

Modern web development frameworks, such as Spring~\cite{Spring} and Django~\cite{Django}, are endeavoring to streamline cache management. Many studies~\cite{Mertz2018, Study_Application} explore functionalities that can autonomously identify procedures suitable for caching, aiming to simplify the caching process. However, existing solutions still require developers to explicitly manage cache invalidation~\cite{Survey_Application}, which is crucial for maintaining data freshness and preventing application errors. Unfortunately, this introduces complexity to the development process.

Existing practices for cache invalidation predominantly incorporate two approaches. The first approach relies on timers, whereby developers actively set a Time-To-Live (TTL) limit for each (type of) cache entry. An entry is automatically discarded when its TTL expires. However, determining an appropriate TTL proves challenging for developers~\cite{Quaestor}. A TTL that is too short undermines cache hit rates, while a TTL that is excessively long risks generating stale cache entries and instigating application errors.
The second approach employs customized cache invalidation rules, demanding developers to identify precisely which cache entry should be invalidated by which data operation. This approach, while allowing more precise cache invalidation, is heavily dependent on the expertise of developers and could result in over-engineering~\cite{Cachematic}. Furthermore, achieving precise cache invalidation through manually defined rules is not always feasible.

In essence, both methods expose the intricate challenge of cache management to developers, necessitating them to balance data freshness with cache efficiency or intertwine sophisticated cache invalidation rules with business logic. As application architectures continue to grow in complexity~\cite{DBLP:journals/pvldb/LaignerZSLK21}, it is imperative to explore methods that render cache management transparent to developers.

In this paper, firstly, we conducted a comprehensive survey to discern the typical challenging cases in application-level cache management. Although prior surveys have delved into application-level caches~\cite{Study_Application, Survey_Application}, they discuss little the challenges in cache invalidation, a pivotal aspect in enabling transparent cache management. Our study attempts to bridge this gap by analyzing 20 web application projects on GitHub and collecting survey responses from more than 50 experienced software engineers. The results highlighted the significance and complexity of cache invalidation. Moreover,  they indicated the current impracticality of achieving transparent and precise cache invalidation solely at the application level, suggesting the necessity for advancements in cache and database systems.

Next, we introduce our design for transparent validity management of the application-level cache.  It is expected to invalidate cache entries once their results are affected by updates of the source data while avoiding premature invalidation. However, our survey reveals that mapping database updates to specific cache entries poses challenges, especially when dealing with complex queries (e.g., queries with joins, range, or multi-cast predicates). In response, we propose modifying traditional DBMS to provide the necessary information for building the links between cached queries and updates.  Meanwhile, an index is employed on the cache side to utilize that and automatically identify cache entries affected by database updates. 

Based on the proposed design, we further provide two specific solutions. The first one leverages the query predicates to judge whether an updated tuple would change the result of a cached query, while the second one employs the bloom filter to do the matching between the query result and its source data. In both solutions, the index on the cache side plays an important role. Therefore, we provide an efficient index for each solution. Our experiments have confirmed that both solutions could enable transparent cache invalidation and effectively improve cache utility compared to traditional TTL-based approaches.
%method, decreasing it from over 50\%.

The rest of the paper is organized as follows. Section~\ref{Sec_Survey} presents the challenges of cache invalidation revealed by our survey. Section~\ref{Sec_Auto}  formulates the problem and presents a framework for transparent cache invalidation. Section~\ref{Sec_solu} and  \ref{Sec_solu2} introduce two specific solutions of the framework. Then, there is a discussion and comparisons between the two provided solutions in Section~\ref{sec_dis}.
A primary evaluation is reported in Section~\ref{Sec_eval}. We discuss the related work in Section~\ref{sec_relat} and future research opportunities in Section~\ref{sec_conclu}.

\section{Challenges in Application-level cache invalidation} \label{Sec_Survey}

To guide the design of the transparent cache invalidation mechanism, we performed a study on open-source Web applications and an online survey targeting software experts, to answer the following questions:
%1) In which scenarios does precise cache invalidation present challenges? 2)  Do developers perceive precise cache invalidation as significant for optimizing web applications?

%\begin{comment}
\begin{enumerate}
    \item In which scenarios does precise cache invalidation present challenges?
    \item Do developers perceive precise cache invalidation as significant for optimizing web applications?
\end{enumerate}    
%\end{comment}

\subsection{ Analysis of Open-Source Repositories} \label{sec_github}
We first conducted a study on the web application projects on GitHub to understand how application-level caching is used in existing web applications.

\subsubsection{Selection of Repositories}
We first searched on GitHub for web projects using the keywords `spring', `spring-boot', and `web'. 
Out of the top 100 returned projects, 72 adopt application-level cache, among which, 17 projects perform cache management, 
 including defining what and where to cache and when to invalid, using handcrafted code, and 55 rely on libraries of web or frameworks, such as Spring cache\footnote{\url{https://docs.spring.io/spring-framework/docs/4.3.x/spring-framework-reference/html/cache.html}}, MyBatis cache\footnote{\url{https://mybatis.org/mybatis-3/sqlmap-xml.html\#cache}} and Hibernate cache\footnote{\url{https://docs.jboss.org/hibernate/orm/6.2/userguide/html_single/Hibernate_User_Guide.html\#caching}}, to manage caches. 

For ease of study, we selected 20 projects applying Spring cache as the subjects of our study. The project selection is mainly based on code quality and the number of stars on GitHub. These projects are shown in table~\ref{table:app}. They cover a wide variety of application domains, including blogging platforms, student data management systems, health data management systems, etc. The table also shows the storage systems, termed ``Cache Store'', that are utilized to store the cache data, and the ORM frameworks, termed ``ORM'', that are applied to map objects in the application with relational tables. We could see that the Spring cache can cooperate with various cache stores and ORM frameworks, which is also conducive to its widespread utilization.

\begin{table}
	\setlength{\belowcaptionskip}{-0cm}
	\centering
	\caption{Applications for analysis.}
	\label{table:app}
	\begin{small}
		\begin{tabular}{l|l|l|l}
			\toprule
			Application name & ORM &   Cache Store & star \\
			\midrule
			halo$^{\textbf{S1}}$				    	& Spring data   & 	    Ehcache &	29.1k \\
			hsweb$^{\textbf{S2}}$ 				    	& MyBatis       &	 	Others  &	8.2k\\
			eladmin$^{\textbf{S3}}$				    	& Spring data   &   	Redis   &	20.3k \\
			21-points$^{\textbf{S4}}$			    	& Hibernate     &	 	Ehcache &	283\\
			Ffast-Java$^{\textbf{S5}}$			    	& MyBatis       &	 	Redis   &	104\\
			FlyCms$^{\textbf{S6}}$  			    	& MyBatis       &	 	Ehcache &	596\\
			iBase4J$^{\textbf{S7}}$				    	& MyBatis       &	 	Redis   &   1.6k\\
			layIM$^{\textbf{S8}}$				    	& MyBatis       &	 	Redis   &	256\\
			sample-boot-hibernate$^{\textbf{S9}}$   	& Hibernate     &   	Ehcache &	105    \\
			dokit$^{\textbf{S10}}$				    	& Spring data   &   	Others  &	486\\
			DimpleBlog$^{\textbf{S11}}$			    	& MyBatis       &	 	Others  &	536\\
			DouBiNovel$^{\textbf{S12}}$                 & MyBatis       &	 	Redis   &	107\\
			xboot$^{\textbf{S13}}$				    	& MyBatis       &	 	Redis   &	3.7k\\
			jeecg-boot$^{\textbf{S14}}$			    	& MyBatis       &	 	Redis   &	29k\\
			javaQuarkBBS$^{\textbf{S15}}$	        	& Hibernate     &	 	Ehcache &	869\\
			jcalaBlog$^{\textbf{S16}}$				    & MyBatis       &	 	Ehcache &	651\\
			Microservices 	platform$^{\textbf{S17}}$	& MyBatis       &	 	Redis   &	4.3k\\
			meetingfilm$^{\textbf{S18}}$			    & MyBatis       &	 	Ehcache &   217\\
			SpringBlog$^{\textbf{S19}}$			    	& Hibernate     &	 	Ehcache &	1.6k\\
			Guns$^{\textbf{S20}}$					    & MyBatis       &	 	Ehcache &	3.7k\\
			\bottomrule
		\end{tabular}
	\end{small}
\end{table}

\subsubsection{Statistics of Cache Usage} \label{Sec_CacheUse}
%Spring cache supports caching at the method level. 
Spring cache~\cite{Spring} is a widely adopted cache management framework and provides an annotation-based caching mechanism. Developers can specify to cache results of a specific method or query using the annotation $@Cacheable$.
By using the annotation $@CacheEvict$, they can also declare rules to invalidate cache entries.
Usually, a cache entry should be invalidated when a method updates its source data, i.e., by executing a certain INSERT, DELETE, or UPDATE statement over the database.

Specifically, besides TTL, Spring cache supports both fine-grained and coarse-grained cache invalidation strategies~\cite{Study_Application}. In the fine-grained approach, cache entries are linked to specific update methods, requiring meticulous rule engineering by developers. On the other hand, the coarse-grained approach associates cache entries with database tables using namespaces, causing a database update to evict all cache entries in the same namespace. It is evident that while the fine-grained approach demands more effort, the coarse-grained approach may suffer from excessive false invalidation~\cite{Cachematic}.

We counted the number of cached methods in each project that adopt fine-grained cache invalidation, coarse-grained cache invalidation (including TTL), or both.
We found that 80.6\% of cached methods adopt coarse-grained or TTL approaches, while only 10.4\% opt for fine-grained ways. Additionally, 9\% of methods chose both.

The result shows that only a minority of methods adopt fine-grained invalidation while using Spring cache. We hypothesize two possible reasons for this practice: (1) fine-grained invalidation is unnecessary, and coarse-grained invalidation and TTL are sufficient to match the application requirements on cache freshness and hit ratio ; (2) invalidation rules are desirable, but they are burdensome or difficult to define. We investigate further to verify which one is the case and whether it is relevant to develop a better solution for fine-grained invalidation.    
%However, most of the time, programmers would or could only choose a simpler and less error-prone way.

%It can be reasoned that fine-grained invalidation rules will be difficult to compose when the cached method contains complex database queries. Following this hypothesis, 

%We could conclude from these statistics that $M$-Type projects hold more complex queries. As discussed before, cache management of MyBatis applies a coarser way, which is also simpler and less prone to error, and more friendly to complex queries.

%	However, 4 of 10 projects using MyBatis cache cached range queries.
%It is understandable as complex queries are too tough for programmers to judge when should be invalided.
%In other words, fine-grained invalidation mechanism in Spring cache will lose its advantage under such situations. 

\subsubsection{Case Analysis} \label{Sec_Scene}
To delve deeper into the reasons behind the limited adoption of fine-grained invalidation, we conducted a further code review of methods employing  TTL and coarse-grained invalidation. We found that the key challenge lies in how to map database updates to cache entries. The following are some typical cases we found in the surveyed projects.

\textbf{Case 1}: 
%\textit{Complex update methods} that involve batch updates of multiple tuples present challenges in defining fine-grained invalidation rules. 
The following code snippet from the project $S6$ executes 11 SQL statements in a sequence, attempting to update 9 tables. As a result, it is impossible to infer merely at the application level about which tuples are actually updated and how they will affect the entries in the cache. 

\vspace{1mm}
\begin{minipage}{0.92\linewidth}
	\small
	\begin{lstlisting}[language=JAVA, frame=trbl]
public DataVo deleteArticleById(Long id) {
  ...
  articleDao.deleteArticleById(id);
  articleDao.deleteArticleCountById(id);
  articleDao.deleteArticleCommentById(id);
  articleDao.deleteAllArticleVotesById(id);
  articleDao.deleteArticleAndCategoryById(id);
  feedService.deleteUserFeed(article.getUserId(),1,article.getId());
  ...
  return data;
}
	\end{lstlisting}
\end{minipage}

%\textbf{Case 2} \textit{The absence of a clear connection between cached methods and updates}  poses challenges in identifying the relationship between cache entries and update statements. Further, it includes the following two types:

\textbf{Case 2}: 
%Complex queries, including join queries and those with range or multicast predicates,  pose challenges in tracking the relationship between a cache entry and its source data. 
The following is another case that adopts coarse-grained invalidation in the project $S6$. The cached method issues a complex query, involving a number of range and multicast predicates. It is thus difficult for developers to figure out what update will change the results of this query, thus invalidating the corresponding entries in the cache.

\vspace{1mm}
\begin{minipage}{0.92\linewidth}
	\small
	\begin{lstlisting}[language=sql, frame=trbl]
Select count(*) From fly_message
 Where from_id = #{fromId} and to_id = #{toId}
  and subject Like CONCAT(CONCAT('%', #{subject}),'%') 
  and send_time Between STR_TO_DATE(#{sendT},'%Y-%m-%d %H')
  and write_time Between STR_TO_DATE(#{writeT},'%Y-%m-%d %H')
  and has_view = 0 and is_admin = 1 and state = 1;
\end{lstlisting}
\end{minipage}

\textbf{Case 3}:
%Different attributes can be used in cached queries and updates, making it difficult for developers to identify the relationship between them even when both of them are relatively simple. 
The following two methods are extracted from the same project. They each issue a simple query to the database, one for read and one for update. The read method is flagged as cacheable. $\textit{user\_name}$ and $\textit{user\_id}$ are different attributes they use to access the same table. Without knowing the exact contents of the table, it is infeasible to infer which $\textit{user\_name}$ is associated with which $\textit{user\_id}$. Therefore, a fine-grained invalidation rule cannot be defined.

\vspace{1mm}
\begin{minipage}{0.92\linewidth}
	\small
	\begin{lstlisting}[language=JAVA, frame=trbl]
public User findByUsername(String userName) {
  return userDao.findByUsername(userName);
}

public int updateAvatar(Long userId,String avatar) {
  return userDao.updateAvatar(userId,avatar);
}
	\end{lstlisting}
\end{minipage}

\begin{minipage}{0.92\linewidth}

	\begin{lstlisting}[language=sql, frame=trbl]
findByUsername:
select * from fly_user where user_name=#{userName} limit 1;
  
updateAvatar:
update fly_user set avatar=#{avatar} where user_id=#{userId};
	\end{lstlisting}	 
\end{minipage}

All the aforementioned cases converge on a single fundamental challenge: the absence of a link between database updates and cache entries at the application level, making precise cache invalidation difficult to achieve.

\begin{figure}[!t]
	
 \centering
 \includegraphics[width=0.85\linewidth]{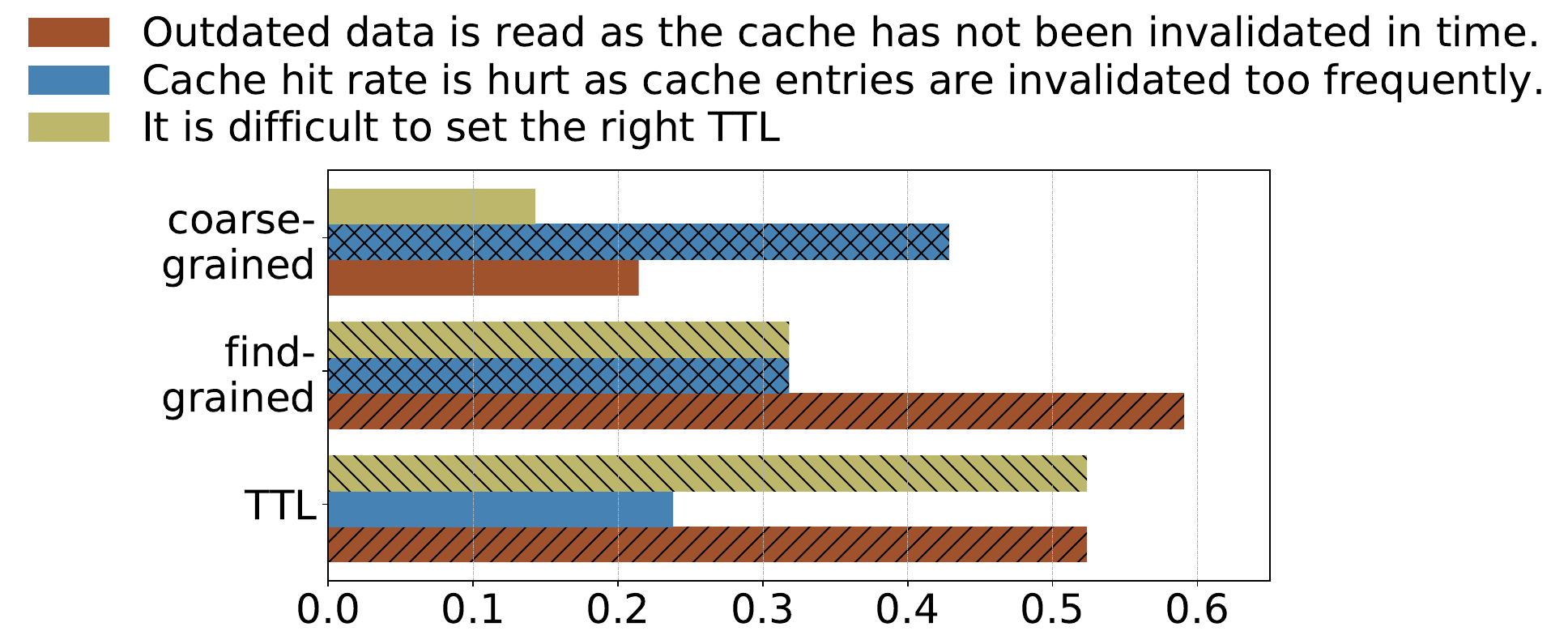}
	\caption{ Percentage of participants bothered by different issues, classified by their preferred invalidation granularity. }
	\label{invalidationStrategy}
\end{figure}

\subsection{Feedback from Online Survey}
To further validate the relevance of researching better cache invalidation methods, we conducted an online expert survey. The questionnaire was distributed to developers of open-source projects through SurveyMonkey\footnote{\url{https://www.surveymonkey.com/r/QVB8CQJ}} and WenJuanXing\footnote{\url{https://www.wjx.cn/vj/to3Cp3T.aspx}}, resulting in 64 high-quality responses.  The participants included 40 software developers, 13 architects, 7 maintenance engineers, and 4 professionals with other occupations.

In the survey, we asked about the issues that bothered the participants most often when they used cache.
Figure~\ref{invalidationStrategy} breaks down the issues across the groups of participants who prefer different invalidation strategies. We can see that different strategies encounter different issues.
As expected, those choosing coarse-grained invalidations suffer more from a lower cache hit ratio. In contrast, those choosing TTL (which is coarse-grained) have trouble with setting an appropriate TTL value. Besides, outdated cached data gains much attention, especially for those applying TTL and fine-grained invalidation strategies.

\begin{table}
	\setlength{\belowcaptionskip}{0cm}
	\setlength\tabcolsep{0.5pt}
	\centering
	\caption{\small What may prevent you from adopting fine-grained invalidation?}
	\label{table:reason}
 %	\resizebox{\linewidth}{!}{
 \small

		\begin{tabular}{l|l}
			\toprule
			Reason & \# \\
			\midrule
			R1: It complicates programs; the many-to-many relation between					& \\
			 tuples and cache entries is difficult to sort out.					& 28 \\
			\midrule
			R2: Queries are too complex to reverse engineer, making it	impossible						& \\
			  to infer which cache entries to invalidate.			& 24 \\
			\midrule
			R3: Update statements are complex, making it impossible	to know			& \\
			 exactly which rows are affected.										& 23 \\
			\midrule
			R4: Different predicates are used by queries and update	statements,			& \\
			making it impossible to know their relationships.				& 11 \\
			\midrule
			R5: Fine-grained invalidation is unnecessary, as it doesn't bring benefits.		& 7 \\
			
			\bottomrule
		\end{tabular}
  %	}

\end{table}

In theory, fine-grained invalidation is superior to coarse-grained invalidation and TTL, as it is more precise and can reduce false invalidation (we have proved it in Section~\ref{Sec_expOverhead}). 
Therefore, we asked participants about the potential reasons that hinder their adoption of finer-grained cache invalidation. The votes from the participants are summarized in Figure~\ref{table:reason}. The majority of participants recognize the benefits of precise cache invalidation, with only a minority considering it unnecessary. However, most participants expressed difficulties in creating and maintaining precise invalidation rules. The issues they pointed out are consistent with the results of our code review in Section~\ref{Sec_Scene}.

Moreover, participants were surveyed regarding the types of database queries for which they typically apply caching, including point queries, range queries, join queries, and multi-cast queries. Their support rates are 60.9\%, 53.1\%, 46.9\%, and 31.2\%, respectively, which indicates their importance in the application-level cache.

Overall, the survey results indicate that fine-grained cache invalidation is perceived as valuable by participants, but challenges in maintaining cache freshness hinder its widespread adoption.

\section{Transparent Cache Invalidation} \label{Sec_Auto}
In this section, we provide a general framework for transparent cache invalidation at first. Then we provide a theoretical analysis of its performance compared to other invalidation strategies.
%Before presenting the detailed design, let's first formulate the question regarding the invalidation of the application-level cache. 
 \subsection{The Framework}
\begin{figure}[!t]
	\centering
	\includegraphics[width=0.8\textwidth]{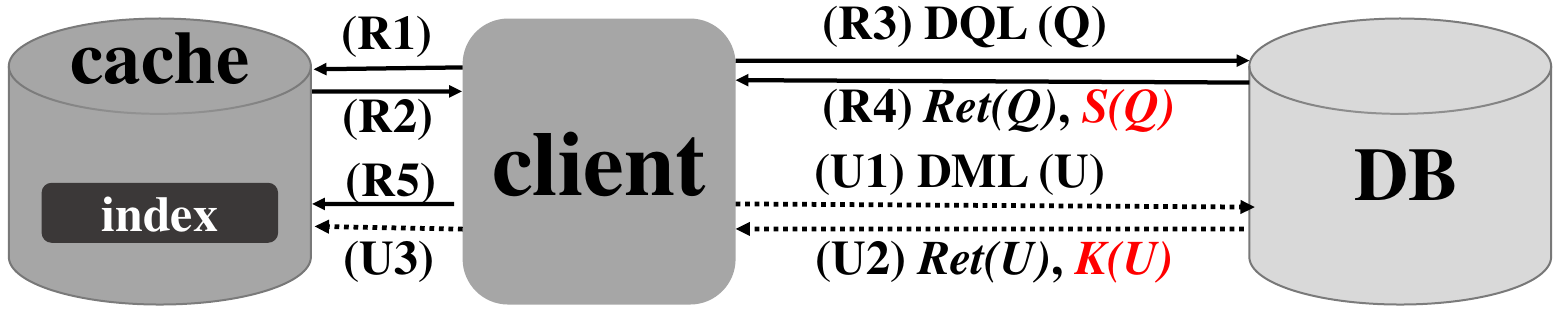}
	\caption{\small The procedure of read(R)/update(U) methods, where \textit{Ret} is the response from the database system, \textit{S(.)} is the signature of DQL statements, and \textit{K(.)} is the signature  of DML statements.}
	\label{framework}
\end{figure}

Figure~\ref{framework} depicts a typical architecture for applying application-level cache, consisting of a client interacting with the application, a database system storing the data, and a cache store storing results of frequently invoked application methods. When conducting read requests, the client first checks the cache~{(R1)}. If it contains the results, they are directly returned~{(R2)}. Otherwise, the corresponding method is executed, and DQL statements (i.e., SELECT statements) are directed to the database~{(R3)}. Once the results are obtained from the database~{(R4)}, they are packaged into a cache entry and added to the cache~{(R5)}. 
When conducting update requests, the client submits a set of DML statements (including UPDATE, INSERT, or DELETE statements) to the database~{(U1, U2)}. Simultaneously, the application is supposed to invalidate the cache entries\footnote{Here, eventual consistency is assumed. Otherwise, the database should wait for the invalidation to complete before committing.} affected by the update~{(U3)}.  However, according to our previous analysis,  achieving precise cache invalidation in such cases is not always possible because the links between cached queries and updated data are unknown to the client. 

Our framework of transparent cache invalidation aims to re-establish the links between cache entries and database updates so that cache invalidation can be done in a precise and automatic way. To achieve this, we propose to make some incremental modifications to both the database system and the cache system.

\textbf{Modification on the database side.} We propose to modify the database system so that it returns additional information after executing a DQL or a DML statement. In particular, the following principles hold for the database system.
\begin{enumerate}
\item When executing a DQL statement $Q$, the database generates a signature $S(Q)$ and returns $S(Q)$ along with the query results, $\mathit{Ret}(Q)$, to the client. In particular, {the signature can be encapsulated into an extension part in the data structure of query results and transparent to those without needing it.} 
\item Similarly, when the database is updated by a DML statement $U$, it will generate a signature $K(U)$  as an extension along with the response $\mathit{Ret}(U)$ to be sent back to the client. 
\item The relationship between $Q(.)$ and $U(.)$ can be defined by a function $F$, such that: 1) if $\mathit{Ret}(Q)$ is modified by $U$, then $F( S(Q), K(U) ) = \mathit{True}$; 2) otherwise, if $\mathit{Ret}(Q)$ is intact, then it is high likely that $F( S(Q), K(U) ) = \mathit{False}$.
\end{enumerate}
As we can see, the signatures $S(.)$ and $K(.)$, together with the function $F(.)$, allow us to establish the links between query results and updates. For each update, we can evaluate its signature against those of previous queries on $F(.)$ to see if it will modify their results.

\textbf{Modification on the cache side.}  An index is maintained by the cache system, through which we can identify all $S(Q)$s that satisfy $F( S(Q), K(U) ) =  \mathit{True}$ for a given $K(U)$. As the signatures are linked to the cache entries, this allows the cache to quickly identify the cache entries affected by the update $U$. 

In this framework, the design of functions  $S(.)$, $K(.)$, and $F(.)$ will be essential and can have different specific implementations. 
In Section~\ref{Sec_solu} and \ref{Sec_solu2}, we present two different solutions of them.

\subsection{Performance Analysis} \label{sec_perAnal}
To gain a theoretical understanding of the gain and loss of transparent invalidation strategy, we consider a simple situation where the cache can accommodate $C$ entries and $u$ DML statements are issued to the database per unit of time. We assume that the cost of executing a database update is $c_u$ and that of refilling a cache entry is $c_q$. When applying transparent invalidation, additional overheads are introduced, increasing to $c_u'$ and $c_q'$, respectively.  Additionally, we assume that, on average, each update outdates $n$ cache entries. We measure the overall cost as the cost of each update plus the cost it incurs to refill the cache. With these assumptions, the costs of handling each DML statement associated with the transparent invalidation, coarser-grained invalidation, and TTL-based invalidation strategies can be calculated as follows, where $p'$ and $p$ are the false positive rates, i.e., the possibility of entry being falsely invalidated, and $t$ is the expiration time.
\begin{itemize}
\item transparent: $(n + (C-n) \times p')\times c_q'+ c_u'$.
\item coarse: ${ (n + (C-n) \times p)\times c_q } + c_u$. 
\item TTL-$t$: $ \frac{C\times c_q}{t\times u} + c_u$.
\end{itemize}

1Generally, the false positive rate $p'$ is expected to be near 0 for transparent invalidation strategies. Therefore, the cost of transparent invalidation primarily arises from the additional overheads of signature generation and maintenance of the invalidation index, quantified as $c_u'$ and $c_q'$, respectively. In our experiments in Section~\ref{Sec_eval}, we found that $c_u'$ and $c_q'$ were at most a few times higher than $c_u$ and $c_q$. On the other hand, the cost of coarse invalidation stems from false invalidation. When the granularity remains at the table level, the false invalidation rate ($p$) may be extremely high, resulting in significant costs. As for the cost of using the TTL approach, it mainly depends on the timer setting ($t$). If the application can tolerate stale data, it can set a relatively large $t$ to reduce the cost. However, for applications sensitive to stale data, using a small $t$ can significantly increase the cost, as confirmed by our experimental results.

%In conclusion, transparent cache invalidation is expected to enhance cache utility, leading to overall performance improvements. While it may introduce certain overheads, the automated and precise cache management offered by transparent cache invalidation can justify these costs. Particularly for applications prioritizing data freshness, transparent cache invalidation emerges as a highly recommended solution.

\section{Solution \uppercase\expandafter{\romannumeral1} -- with Predicates} \label{Sec_solu}

\subsection{The Design} \label{sec_predic}
In this part, we assume that the DQL queries to the database are limited to conjunctive SPJ (i.e., Select, Project, and Join) queries, where the selection predicates include only point, range, and multicast predicates. Our survey on GitHub revealed that the vast majority of queries issued to databases belong to these types. Moreover, as disjunctive queries can usually be decomposed as a union of conjunctive queries, this design should directly apply to most cases of disjunctive queries too.

In the first solution, we utilize the predicates of DQL queries as their signatures. In particular, we first consider single table queries, and the functions of $S(.)$, $K(.)$, and $F(.)$ are defined as follows:
\begin{itemize}
\item $S(Q) \triangleq  \{ (a, \mathcal{P}) | \mathcal{P} \mbox{ is a predicate of } Q  \mbox{ on attribute }a \} $
\item $K(U)\triangleq  \{t | t \mbox{ is a tuple updated by } U \}$
\item $F(S(Q), K(U)) \triangleq  \bigvee_{t\in K(U)} (\bigwedge_{(a, \mathcal{P})\in S(Q)} \mathcal{P}(t.a))$
\end{itemize}
In other words, $S(.)$ returns all the predicates of the query\footnote{Here, we assume each predicate only utilizes one attribute. Predicates utilizing multiple attributes can be regarded as based on a virtual attribute. For example, the predicate $R.a1 > R.a2$ could be regarded as based on the virtual attribute $R.a3=R.a1 - R.a2$ and transformed into $R.a3>0$.}. $K(.)$ returns the set of updated tuples. And $F(.)$ evaluates whether there is at least one tuple in $K(.)$ satisfies all the predicates in $S(.)$. 

For example, given the following query $Q1$, its signature will be the predicate $S(Q1) = \{(R1.a1, \mbox{ between }C1\mbox{ and }C2)\}$. If a DML query $U1$ updates the table $t_1$ in $R1$, we can generate the signature $K(U1) = \{t_1\}$. In $F(S(Q1), K(U1))$, we evaluate $t_1.a1$ against the predicate, to determine if $U1$ modifies the results of $Q1$.  
In particular, sending the whole updated tuple back to the client may be expensive. Indeed, only attributes utilized by DQL query predicates are needed. Therefore, the database can memorize queried attributes and only includes those attributes of updated tuples in the signature. For the above example, only $t_1.a1$ is needed and $K(U1)$ can be $\{t_1.a1\}$

\vspace{1mm}
\begin{minipage}{0.92\linewidth}
	\small
	\begin{lstlisting}[language=sql, frame=trbl]
Q1: Select a1,a2,...,aN From R3  Where R1.a1 between C1 and C2; \end{lstlisting}  
\end{minipage} 

When a DQL query involves joins over multiple tables, the above design must be adjusted to work. For example, the following query $Q2$ contains a join on tables $R2$ and $R3$ and a selection predicate only on $R3$. When a tuple in $R2$ is updated, it may change the results of $Q2$. However, this case cannot be detected by the above design. %its signature is defined as follows, where $R1 \bowtie R2$ is the virtual joined table.

\vspace{1mm}
\begin{minipage}{0.92\linewidth}
	\small
	\begin{lstlisting}[language=sql, frame=trbl]
Q2: Select * From R2, R3 
     Where R2.foregin_key = R3.primary_key and R3.a2 between C1 and C2;
	\end{lstlisting}  
\end{minipage}

To handle such a case, we extend the design of $K(.)$.
In particular, after processing $Q2$, the database memorizes its template, which truncates all the selection and projection operators and preserves only joins. %to facilitate the retrieval of necessary buddy tuples from another source table during tuple updates. 

\vspace{1mm}
\begin{minipage}{0.92\linewidth}
	\small
	\begin{lstlisting}[language=sql, frame=trbl]
Template(Q2): Select * From R2, R3 Where R2.foregin_key = R3.primary_key;
	\end{lstlisting}  
\end{minipage}

\noindent In case a tuple $t_2$ in table $R2$ is updated by a DML statement $U2$, the database replaces $R2$ in the template with $t_2$, resulting in the following query.

\vspace{1mm}
\begin{minipage}{0.92\linewidth}
	\small
	\begin{lstlisting}[language=sql, frame=trbl]
Q3: Select * From %*$t_2$*), R3  Where  %*$t_2$*).foregin_key = R3.primary_key;
	\end{lstlisting}  
\end{minipage}

\noindent This query will link $t_2$ to a tuple $t_3$ in $R3$, and return a joint tuple $(t_2,t_3)$ as the result. This joint tuple will then be returned as the signature of $U2$, which can be evaluated against the predicate on $R3$ to determine if the update will affect $Q2$. 

In simple terms, when the database receives a DQL query flagged as cacheable, it stores its query template in its memory. Subsequently, upon receiving a DML statement $U$, the database uses the updated tuple to instantiate the memorized query templates to generate a series of queries (in the form of $Q3$). Then, the database executes the queries to generate a set of joint tuples, which serve as the signature $K(U)$. %To accelerate these queries, an index on the join attributes is advisable. 

Undoubtedly, this process will increase the overhead on the database when handling updates. However, we believe such overheads remain manageable for most real-world web applications. First, they typically involve only a limited number of query templates within their programs. Based on our survey of GitHub projects, each table in an application is linked to an average of 0.34 and a maximum of 2 join query templates. Second, our survey also shows that each join query, on average, involves 2.2 tables. It means that, in most cases, the query generated using the above method will be a simple selection query (in the form of $Q3$), which can be accelerated using indexes on the join attributes.

\subsection{The Index for Predicates} \label{Sec_Index}

On the cache side, we face the challenge of designing an index capable of identifying $S(Q)$ that satisfies $F(S(Q), K(U)) = True$ for any given $K(U)$.  In line with our design, the index should be adept at identifying predicates (or predicate groups) fulfilled by an updated (joint) tuple. While various approaches exist for indexing individual predicates of the same type, indexing groups of predicates with different types proves to be a nontrivial task.

%The invalidation index on the cache side, designed to handle point, range, and multicast queries uniformly, serves as a versatile index for identifying cache entries affected by DML statements. 
In this solution, we utilize Q-Tree, which is a variant of the interval tree~\cite{Algorithms} and will be further introduced in Section~\ref{Sec_qtree}, as the index and transform predicates of different types in $S(Q)$ all into intervals. In the following, we enumerate how to perform the transformation for point, range, and multicast predicates.

A range predicate essentially represents an interval. For instance, consider the predicate in the previous query $Q1$, which can be transformed into the interval $\mathcal{I}=[C1, C2]$ on the attribute $\textit{R1.a1}$. Suppose we index this interval using an interval tree. When a tuple $t$ in $R1$ gets updated, we can search the interval tree to retrieve all the intervals encompassing the value of $t$ on $a1$. If the search results include $\mathcal{I}$, it indicates that the update will modify the results of $Q1$.

Point predicates (i.e., the exact-match selection) can be regarded as special intervals whose lower and upper bounds are identical. 

A substring multicast predicate can also be transformed into an interval. For example, consider the following query $Q4$, which contains a multicast predicate specifying that the attribute ${R4.a2}$ should contain a substring $'\textit{hot}'$. We can represent it as an interval $['hot', 'hot\#']$, where $'\#'$ represents the largest literal and $'hot\#'$ represents the upper bound of strings prefixed with $'hot'$.

\vspace{0.5mm}
\begin{minipage}{0.92\linewidth}
	\small
	\begin{lstlisting}[language=sql, frame=trbl]
Q4: Select * From R4 Where R4.a2 like '%hot%';
	\end{lstlisting}  
\end{minipage}

\noindent During an update, the string of the updated tuple needs to be transformed into multiple suffix strings for evaluation against the interval. 
For instance, if the tuple $t_4$ with ${t_4.a2}$ = $'So~hot!'$ is updated,  all the suffixes of ${t_4.a2}$ should be evaluated against $['hot', 'hot\#']$, including $'So~hot!'$, $'o~hot!'$ and so on. Eventually, the suffix $'hot!'$ matches the interval $['hot', 'hot\#']$, indicating that the update will indeed modify the results of $Q4$.

In our enhanced cache system, an interval tree is created for each attribute in the database used for selection. For every signature $S(Q)$, which is a group of predicates, a deterministic procedure is employed to identify the most selective predicate within $S(Q)$. This predicate is then transformed into an interval and inserted into the appropriate interval tree.
During an update, we evaluate each tuple in $K(U)$ against all relevant interval trees to retrieve matching signatures. We then double-check if the predicates in each signature are all satisfied by the tuple. This guarantees to identify all $S(Q)$ that satisfies $F(S(Q), K(U)) = True$.

There is a possibility that none of the predicates in a signature is individually selective enough to ensure the precision of indexing. In such cases, we can utilize the Hilbert Curve~\cite{Hilbert} to transform multiple predicates on different attributes into a single-dimensional interval~\cite{HilbertSplit}. Then, we index this transformed interval to achieve improved precision.

%\subsection{Overheads}\label{Sec_Perf}
%To achieve transparent cache invalidation, our solution introduces overheads to both the cache side and  the database side.

Indexes on the cache side can contribute to storage and lookup overhead. Their costs are related to the number of interval trees, which further depends on the number of attributes used in database selection. Nevertheless, according to some empirical studies~\cite{AOP2006, qTemplate, Apollo}, most applications typically have a limited number of query attributes. In our surveyed GitHub projects, there are usually fewer than 10 per table.  Therefore, the cost of maintaining invalidation indexes should remain manageable.

%The detector plays a critical role in reporting updated tuples to the cache server. For join queries, it also needs to retrieve buddy tuples from related source tables. The overhead in this case is influenced by the number of buddy tuples retrieved for each updated tuple. However, this paper focuses on scenarios where an updated tuple has a limited number of buddy tuples. Such cases are common in web applications, while scenarios with exploding joins and lots of buddy tuples are more relevant to analytical queries and beyond the scope of this study.

\subsection{Q-Tree}\label{Sec_qtree}
%\subsubsection{Overview}
Q-Tree is the index designed to infer query predicates from a given tuple. 
As mentioned earlier, the items to be indexed in a Q-Tree are simple intervals, represented as a lower bound and an upper bound, which are transformed from the query predicates.

However, when applied to cache invalidation, Q-Tree faces workloads that are drastically different from that of a general-purpose index, such as B+~Tree or Binary Search Trees. In particular, for an index structure in a database system, the majority of workloads are point and range queries. Insertion and deletion are usually less frequent. For a Q-Tree, however, the majority of the workload is insertion and deletion.

In particular, Q-Tree faces 3 main types of workload in cache invalidation: 
\begin{enumerate}
	\item \textbf{insertion}. When an entry is added to a cache, a set of intervals from its predicates will be inserted into the Q-Tree of its related attribute.
	\item  \textbf{eviction}. When a cache entry is evicted because of the fulling of the cache, the corresponding intervals will be deleted from the Q-Tree.
	\item  \textbf{invalidation}. When an update is performed on the database, the cache server will look up the Q-Tree for predicates to invalidate. If a matching predicate is found, a follow-up eviction will be performed.
\end{enumerate}

Specifically, the original Interval Tree~\cite{Algorithms} is a binary tree, which is not suitable for large-scale datasets and highly concurrent situations. Therefore, we extend the interval tree to a B+~Tree-like structure, so that it can work as a balanced n-ary tree and finer-grained locks can be applied to make it more scalable.    

\begin{figure}[htp]
	\centering
	\includegraphics[width=0.88\textwidth]{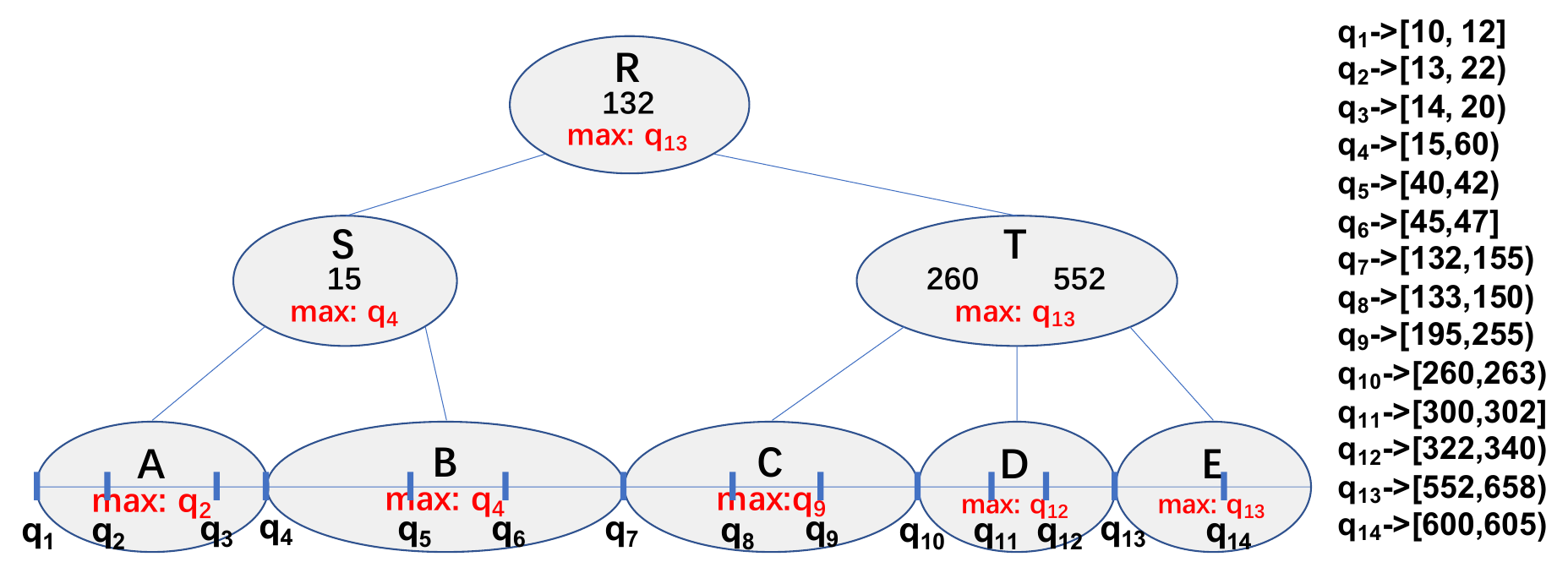}
	\caption{Illustration of a 3-ray Q-Tree.}
	\label{Q-Tree}
\end{figure}

A 3-ary Q-Tree is illustrated in Figure~\ref{Q-Tree}. Physically, it is a B+~tree, whose index keys are the lower bounds of intervals. 
Additionally, each node $X$ maintains $X.max$, which is the upper bound of intervals stored in its subtree.
$X.max$ is employed by the interval tree to locate matching intervals.
During a search, we traverse the Q-Tree using the B+~tree algorithm. As the index keys are the lower bounds of the intervals, the original B+~tree algorithm only allows us to find intervals whose lower bounds are smaller than the search key. We still need to filter out the intervals whose upper bounds are smaller than the search key. This is where $X.max$ comes into play. The concrete search algorithm is similar to that of Interval Tree~\cite{Algorithms}. We further discuss it in Section~\ref{Sec_invalid}.

\begin{algorithm} 

	\SetKwData{key}{key}\SetKwData{querySet}{querySet}\SetKwData{root}{root} \SetKwFunction{cas}{cas}
	\SetKwFunction{unLock}{unLock}
	\SetKwFunction{isLeaf}{isLeaf}
	\SetKwFunction{cover}{cover}
	\SetKwFunction{add}{add}
	\SetKwFunction{remove}{remove}
	\SetKwFunction{addWriteLock}{addWriteLock}
	\SetKwFunction{IsUnderflow}{IsUnderflow}
	\SetKwFunction{recalculateBoundary}{recalculateBoundary}
	\SetKwFunction{remove}{remove}
	\SetKwFunction{remove}{remove}
	\SetKwInOut{Input}{input}
	\SetKwInOut{Output}{output}
	\SetKwFunction{QueryRetrieval}{QueryRetrieval}
	\SetKwFunction{Travel}{Travel}
	\SetKwData{node}{node}\SetKwData{query}{query}\SetKwData{child}{child}
	\SetKwData{childs}{childs}\SetKwData{rebalancedChildNum}{rebalancedChildNum}\SetKwData{formatBit}{formatBit}
 \SetKwData{lock}{lock}
	\SetKwData{modifyLock}{modifyLock}\SetKwData{mergeLock}{mergeLock}\SetKwData{True}{True}
	\SetKwData{0}{0}\SetKwData{1}{1}
	\SetKwData{queries}{queries}
	
	\SetKwProg{Fn}{Function}{:}{}
	\caption{Retrieve and drop queries whose results are related to the given data utilizing Q-Tree}  \label{Algo:invalid}
	\Input{\key: the invalidation key;}
	\Output{\querySet: queries whose ranges cover \key}
	\small
	\Fn{ \QueryRetrieval{\key}}{
		\querySet = $\{\}$ \;
		\Travel {\root, \key, \querySet} \;
		\KwRet \querySet \;
	}
	
	\BlankLine 
	\Fn{\Travel{\node, \key, \querySet}}{
		\eIf{ \node.\isLeaf{}}{
			\For{\query in \node.\queries} {
				\If{ \query.\cover{\key}}{
					\querySet.\add{\query}\;
					\node.\remove{\query}\;
				}
			}
		} { \tcp*[h]{ internal node }\label{lelse} \\
			\For{\child in \node.\childs}{
				\If{ \child.\cover{\key}}{
					\Travel{\child, \key, \querySet}\;}
			}
			
			\If {\node.\rebalancedChildNum $>$ \0  and \node.\formatBit.\cas{\0,\1} == \True}{
				\node.\lock.\addWriteLock{}\;
				\For{\child in \node.\childs}{
					\If{ \child.\IsUnderflow{}}{
						apply write lock for used nodes \;
						do re-balance\;
						unlock  used nodes \;
					}
				}
				\node.\lock.\unLock{}\;
				\node.\formatBit.\cas{\1,\0}\;
			}
		}  
		\node.\recalculateBoundary{}\;
	}
\end{algorithm}

\subsubsection{Invalidation Operation}\label{Sec_invalid}
\textbf{Insertion} and \textbf{eviction} on Q-Trees are similar to that on B+~Tree. Therefore, we put our point on \textbf{invalidation}. 
Specifically, we adopt B-link~Tree~\cite{blink}, which have been adopted by many state-of-the-art database systems\footnote{\url{https://github.com/postgres/postgres/blob/master/src/backend/access/nbtree/README}} to improve the concurrency of B+~Tree.
B-link~Tree complements each internal node of B+~Tree with additional links that point to its sibling nodes. These links glue the broken structures of B+~Tree together during the split or merge process so that we can substantially shorten lock duration and thus increase concurrency.

Algorithm~\ref{Algo:invalid} shows the procedure of how to find and drop queries based on a given key.
Specifically,  similar to B+~Tree, rebalancing is the process of making sure the fanouts of all nodes (except the root) are not below a certain threshold. 
It is usually the heaviest process in B+~Tree's operations.
In the algorithm, we try to let as less threads be blocked to wait to do the rebalancing as possible, as it is time-consuming and exclusive.

In particular, one invalidation operation may invalidate multiple cache entries and then cause multiple droppings in the index.
It is very possible that a node and its siblings both need to be rebalanced, successively.
Then, instead of doing the rebalance immediately after finishing searching on a node, we check nodes' fanouts and do the rebalance after finishing searching on all siblings ($Line~16 \sim~21$). 
In particular, before a node is merged and deleted, an exclusive lock will be applied to it to ensure no other threads access it ($Line~19$).
Meanwhile, multiple concurrent invalidation operations may access the same node and find its child needs to be rebalanced.
In fact, only one thread is needed to perform the job.
A $\mathit{formatBit}$ is used to prevent this wasteful contention.
Atomic operations\footnote{\url{https://gcc.gnu.org/onlinedocs/gcc-4.1.0/gcc/Atomic-Builtins.html}} ($cas$) on $\mathit{formatBit}$ ($Line~15$ and $Line~23$) could make sure only one node would do the rebalance for its children.

\subsubsection{Performance analysis}
Q-Tree has a similar structure as B+~tree and each node is enhanced with an extra $max$ value and $\mathit{mergeLock}$, whose space overhead is negligible.
Therefore, its space complexity is the same as B+~tree, while its retrieving cost is the same as that of the interval trees, which is $O(k*log(N))$~\cite{Algorithms}, where $k$ is the number of intervals that the key falls in and $N$ is the total number of cache entries in the Q-Tree. However, given the situation of cache invalidation, cache entries related to chosen intervals need to be invalided. As a result, those intervals need to be deleted from the index. Therefore, the amortized cost of Q-Tree is $O(log(N))$ \cite{Algorithms}.
%The number of nodes in the Q-Tree is just the same as that of the B+~Tree.
%And 

\section{Solution \uppercase\expandafter{\romannumeral2} -- with Bloom Filters} \label{Sec_solu2}
In the first solution, using predicates as the signature of DQL queries may require intricate coordination between the cache and the database. The cache needs to analyze the predicate to determine how to construct the index effectively. To address this challenge, we propose an alternative solution that leverages bloom filters as the signature to reestablish the connection between cache entries and the source data.

\subsection{The Design}
Intuitively, the matching between a DQL query $Q$ and an updated statement $U$, can be achieved by memorizing the identifiers of all tuples used by $Q$ in its signature $S(Q)$ and including the identifiers of updated tuples in $K(U)$, the signature of $U$. Hence,  the matching between $Q$ and $U$ can be implemented by detecting the intersection between $S(Q)$ and $K(U)$. However, there are two key challenges to consider. First, when a query involves a significant number of tuples, the overhead of maintaining $S(Q)$ can become substantial. Second, in scenarios with a high volume of cache entries, the process of detecting intersections between $K(U)$ and all $S(Q)$s can be computationally expensive. Therefore, a more lightweight design is necessary.

The bloom filter~\cite{BF}, a widely adopted data structure for determining the presence of data in a dataset, offers an elegant solution. It operates with minimal time and space requirements, albeit with the possibility of false positives. As discussed earlier, in the context of cache invalidation, false positive invalidations are tolerable, making the Bloom filter a compelling choice for this purpose.

In the second solution, we employ bloom filters (BFs) to represent the signature of each query. Specifically, we define the functions $S(.)$, $K(.)$, and $F(.)$ as follows:
\begin{itemize}
\item $S(Q) \triangleq  BF_Q, \mbox{ if a tuple } t \mbox{ is accessed by } Q, \mbox{ then } t.pk \in BF_Q$
\item $K(U) \triangleq \{BF_U | t.pk \in BF_U, t \mbox{ is a tuple updated by } U \}$
\item $F(S(Q), K(U)) \triangleq \bigvee_{BF_U\in K(U)} (BF_U \subseteq BF_Q)$
\end{itemize}
In essence, $S(.)$ is a bloom filter holding the primary keys of all tuples accessed by the query, while $K(.)$ is a set of bloom filters where each one contains the primary key of a tuple updated by the DML statement. If the predicate $BF_Q \subseteq BF_U$ is satisfied,  it indicates that the primary key contained in $BF_U$ is highly likely also contained in $BF_Q$. Consequently, the result of query $Q$ is highly likely influenced by the update $U$, necessitating invalidation. If any bloom filter in $K(U)$ satisfies this condition, the cache entry corresponding to $S(Q)$ should be invalidated.

Typically, DML statements have two primary effects on database tables: the deletion of old data and the insertion of new data. In particular, an UPDATE statement can be regarded as a combination of both.
The above solution can effectively handle the case where an old tuple accessed by the DQL query $Q$ is deleted. However, it does not address situations involving inserted tuples. For example, consider the following query $Q5$. When this query is executed, the contents in table $R5$ are depicted in Table~\ref{table:exampleR5}. Consequently, the signature of $Q5$ contains $t_{52}$, $t_{54}$ and $t_{55}$. 

\vspace{0.5mm}
\begin{minipage}{0.92\linewidth}
	\small
	\begin{lstlisting}[language=sql, frame=trbl]
Q5: Select * From R5 Where R5.a1 between 2 and 4;
	\end{lstlisting}  
\end{minipage}

\begin{minipage}[t]{\linewidth}
\begin{minipage}[t]{0.49\linewidth}
	\centering
%\begin{table}
	 \makeatletter\def\@captype{table}\makeatother\caption{Contents in table $R5$.} \label{table:exampleR5}
	\begin{small}
		\begin{tabular}{c|l|l}
			\toprule
			    primary\_key & $a1$& $a2$ \\
			\midrule
			$t_{51}$  & 1 & this is an example \\  
			$t_{52}$  & 2 & this is an example \\  
			$t_{53}$  & 5 & this is an example \\  
   			$t_{54}$  & 4 & this is an example \\ 
         	$t_{55}$  & 3 & this is an example \\  

			\bottomrule
		\end{tabular}
	\end{small}
%\end{table}
\end{minipage}
\begin{minipage}[t]{0.49\linewidth}
	\centering
%\begin{table}
	 \makeatletter\def\@captype{table}\makeatother\caption{Contents in table $R6$.} \label{table:exampleR6}
	\begin{small}
		\begin{tabular}{c|c|l|l}
			\toprule
			    primary\_key & foreign\_key & $a3$ & $a4$ \\
			\midrule
			$t_{61}$  & $t_{51}$ & 1 &text \\  
			$t_{62}$  & $t_{52}$ & 2 &text \\  
   			$t_{63}$  & $t_{51}$ & 3 &text \\  
			$t_{64}$  & $t_{54}$ & 5 &text \\  
                $t_{65}$  & $t_{55}$ & 1 &text \\  
			\bottomrule
		\end{tabular}
	\end{small}
%\end{table}
\end{minipage}
\end{minipage}

\vspace{2mm}

When an update statement $U1$ as follows is executed,  it has an impact on the results of query $Q5$,  necessitating the invalidation of its associated cache entry. However,  a challenge arises because  $t_{51}$ is not contained in $S(Q5)$. Consequently, $F(S(Q5), K(U1))$ cannot identify this change, potentially leading to false negatives, which are unacceptable.

\vspace{0.5mm}
\begin{minipage}{0.92\linewidth}
	\small
	\begin{lstlisting}[language=sql, frame=trbl]
U1: Update R5 Set a1=2 Where primary_key=%*$t_{51}$*);
	\end{lstlisting}  
\end{minipage}

To handle such cases, we extend the design of $K(.)$. In particular, if the result of query $Q$ is cached and a new tuple $t$ could satisfy the predicates of $Q$, there must be a tuple $t'$ near to $t$ and the $t'.pk$ is included in $S(Q)$. Otherwise, the result of $Q$ will be empty and we assume that empty results will not be cached. Considering the above example, even if $t_{51}$ is not in $S(Q5)$, its neighbor $t_{52}$ is contained in $S(Q5)$. Therefore, if $t_{52}$ is included in $K(U1)$, $Q5$ will be identified correctly. 

In our approach, we assume there is an index $I_a$ on the attribute $a$ and define the left neighbor of tuple $t$  with respect to the index $I_a$ as the tuple whose value on attribute $a$ is no greater than and closest to $t.a$. Similarly,  the right neighbor of a tuple $t$  with respect to $I_a$ is the tuple whose value on attribute $a$ is no smaller than and closest to $t.a$.
After processing a query $Q$, the database needs to memorize its utilized indexes, such as the $R5.I_{a1}$ in the case of query $Q5$,  in an index set denoted as $\mathcal{I}$. When generating signatures for insert statements, the database needs to retrieve the primary keys of all left and right neighbors of the new tuple with respect to indexes in $\mathcal{I}$. They are then added to the signatures. Therefore, in the above example, both $t_{52}$ and $t_{54}$ will be included in $K(U1)$, allowing for the accurate identification of $Q5$.

Specifically, the design of  $K(.)$ can be adjusted as follows:
\begin{itemize}
\item $K(U) \triangleq  \{BF_U | t.pk \in BF_U, t \mbox{ is a tuple updated by } U \} $ 
\\
\hspace{2em} $\bigcup \{BF_U | t.pk \in BF_U, \   t \mbox{ is a neighbor of a tuple inserted by } U \}$
\end{itemize}

Furthermore, when multiple tables are involved in a DQL query, the function $S(.)$ requires careful design. Consider the following query $Q6$, and when it is invoked, the contents of table $R5$ and $R6$ are displayed in Table~\ref{table:exampleR5} and Table~\ref{table:exampleR6}, respectively.

\vspace{0.5mm}
\begin{minipage}{0.92\linewidth}
	\small
	\begin{lstlisting}[language=sql, frame=trbl]
Q6: Select * From R5, R6 
    Where R6.foregin_key = R5.primary_key 
      and R5.a1 between 2 and 4 and R6.a3 between 2 and 4;
	\end{lstlisting}  
\end{minipage}

In particular, the result of $Q5$ only contains the joint tuple $\{t_{52}, t_{62}\}$. If its signature $S(Q6)$ only includes $t_{52}$ and $ t_{62}$,  it would lead to false negatives. To illustrate, consider the following UPDATE statement $U2$, which will add the joint tuple $\{t_{54}, t_{64}\}$ into the result of $Q5$. However, the signature of $U2$, which includes $t_{64}$ and $t_{63}$, fails to identify $Q5$ correctly.
This is because even if $t_{63}$ can satisfy the predicate ``$R6.a3$ between 2 and 4'' on table $R6$, it is filtered out because its referenced tuple $t_{51}$ cannot satisfy the predicate on table $R5$. 

\vspace{0.5mm}
\begin{minipage}{0.92\linewidth}
	\small
	\begin{lstlisting}[language=sql, frame=trbl]
U1: Update R6 Set a3=4 Where primary_key=%*$t_{64}$*);
	\end{lstlisting}  
\end{minipage}

To address these cases, the database must retrieve the primary keys of all tuples accessed during the query and include them in the signature. Therefore, $S(Q6)$ includes $t_{52}$, $t_{54}$ and $t_{55}$ from the table $R5$, and $t_{62}$ and $t_{63}$ from the table $R6$\footnote{Assume that proper indexes are established on the attribute $R5.a1$ and $R6.a3$ respectively.}. Consequently, after executing the statement $U2$, $Q5$ can be identified by $t_{63}$.

\subsection{The Index for Bloom Filters}
On the cache side, we need to identify all $S(Q)$s that satisfy $F(S(Q), K(U)) = \mathit{True}$ for a given $K(U)$. In this solution, both $S(Q)$ and $K(U)$ are bloom filters. In other words, we should retrieve all bloom filters that contain a given bloom filter, i.e., all keys inserted into the later one are contained in the former.

In particular, a bloom filter is a data structure consisting of a bit array with $M$ bits (all initiated as 0) and is used in conjunction with $K$ hash functions. When a $key$ is inserted into a bloom filter, it will generate $K$ hash values using these $K$ hash functions,  and the corresponding bits in the bit array are then set to 1. To determine whether a bloom filter contains a given $key$, it checks the bits corresponding to the K hash values generated for that $key$. If all these bits are 1, the key is considered to be contained in the bloom filter.

To efficiently organize the bloom filters generated for various queries, we configure all of them with the same parameters. This means they share the same bit array size and the same set of hash functions. Therefore, given a bloom filter $BF_U$ in $K(U)$, if the bloom filter $BF_Q$ in $S(Q)$ contains $BF_U$, it implies that $BF_Q \& BF_U = BF_U$.

Nevertheless, when dealing with a large number of cache entries, implying numerous bloom filters, the process of detecting whether each bloom filter contains a given one can become prohibitively expensive. As illustrated in Figure~\ref{framework},  a dedicated index, designed to handle bloom filters as its indexed items, becomes essential to efficiently carry out this task.

\begin{figure}[tp]
	\centering
	\includegraphics[width=0.9\textwidth]{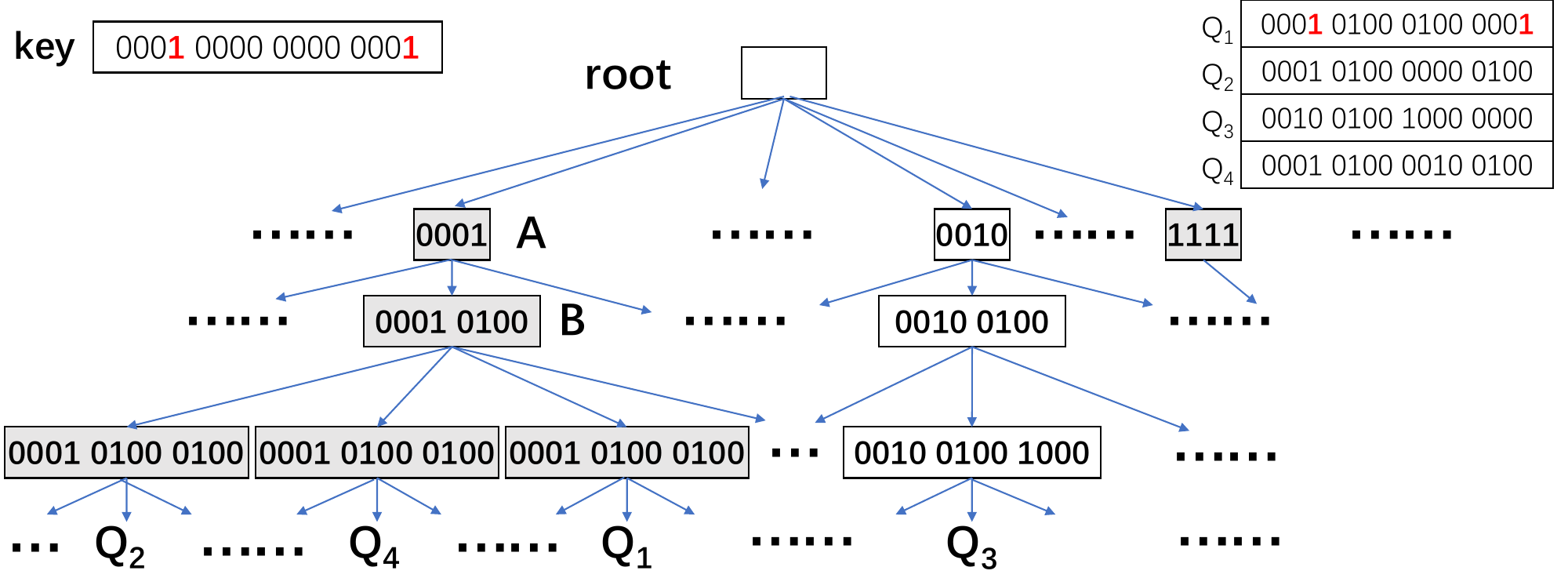}
	\caption{Illustration of a Trie tree for bloom filters with 16 bits.}
	\label{fig:Trie}
\end{figure}

Intuitively, a Trie tree~\cite{algorithms2}, which is a typical data structure for string matching, can be employed for this purpose. However, the bloom filter in $K(U)$ contains just one key, resulting in an extremely sparse bit array with the majority of bits set to 0. This sparsity can lead to inefficiencies when using a Trie tree for matching. For example, as illustrated in Figure~\ref{fig:Trie}, consider a Trie designed for bloom filters with 16 bits.  Each internal node in the tree represents a specific prefix bit array. For instance, node $A$ signifies that the prefix of all bloom filters in its subtree is ``$0001$''. When a key with the bit array ``$0001,0000,0000,0001$'' comes, traversal of all child nodes of node $A$ becomes necessary, as the 5th to 8th bits of the given key are all 0s.

\begin{figure}[tp]
	\centering
	\includegraphics[width=0.9\textwidth]{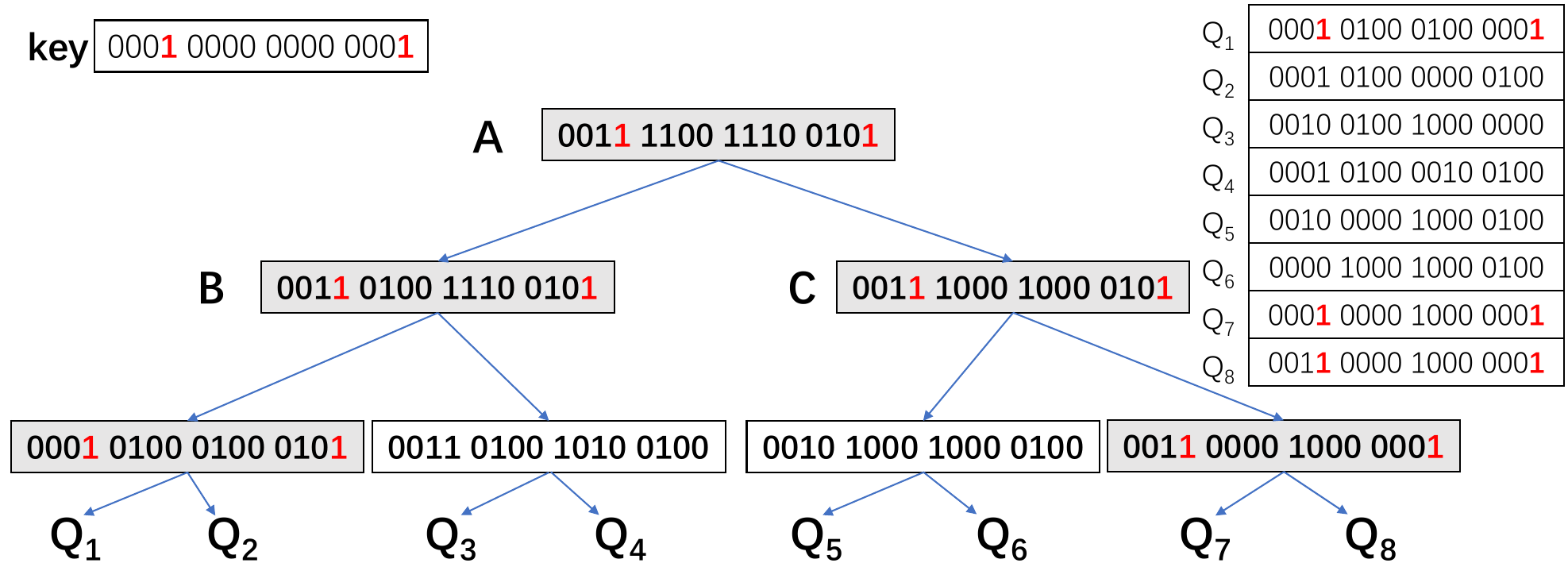}
	\caption{Illustration of a 2-ary BF-Tree for bloom filters with 16 bits.}
	\label{fig:BFTree}
\end{figure}	

In contrast, in this solution, we introduce a BF-Tree to index the bloom filters. Figure~\ref{fig:BFTree} illustrates a 2-ary BF-Tree. In particular, each node in the tree maintains a $\textit{bitMask}$, which is the result of a bitwise $or$ operation across all the bloom filters in its subtree. As indicated in the figure, the $\mathit{bitMask}$ of the root node is the $or$ result of all bloom filters included in the index. 

Specifically, if the $\mathit{bitMask}$ of a node does not contain a given key, it can be confirmed that none of the bloom filters within its subtree includes the given key. Therefore, when searching for a given key, it only needs to traverse nodes whose $\mathit{bitMask}$ could contain the given key. 

Furthermore, it is easy to expand BF-Tree into an $n$-ary balanced tree by applying a similar data structure as Q-Tree.  Consequently, the specific procedure of invalidation operation is akin to the one outlined in Algorithm~\ref{Algo:invalid}, with the condition to determining whether a node covers a key as $\mathit{bitMask} \ \& \ \mathit{key} =  \mathit{key}$.

\section{Comparison between the Two Solutions} \label{sec_dis}

%In this section, we provide a comparison between the above two different solutions.
In the following, we will use the term ``transparent-$\mathcal{P}$'' to denote the first solution introduced in Section~\ref{Sec_solu}, and ``transparent-$\mathcal{B}$'' to represent the second solution introduced in Section~\ref{Sec_solu2}. In this section, we aim to provide a comprehensive comparison between transparent-$\mathcal{P}$ and transparent-$\mathcal{B}$ regarding their ease of implementation and performance.

\subsection{Ease of Implementation}
Both solutions share the same foundational design introduced in Section~\ref{Sec_Auto}, necessitating modifications on both the cache and database sides. However, they diverge in the manner in which they handle data interaction.

In the case of transparent-$\mathcal{P}$, query predicates and utilized attributes of updated tuples are transmitted from the database to the cache. Subsequently, the cache is tasked with constructing the index based on these predicates. While this approach benefits from the availability of query predicates from the query parser of databases~\cite{UrWeb}, it still requires the cache to analyze these query predicates. As discussed in Section~\ref{Sec_Index}, distinguishing between different types of predicates is essential for the proper construction of Q-Trees, which places a high degree of coordination and cooperation between the cache and the database.

In contrast, in the transparent-$\mathcal{B}$ solution,  the data transmitted from the database to the cache consists solely of bloom filters. This approach eliminates the need for caches to comprehend the origin of the query result and the mechanics behind the generation of bloom filters, effectively decoupling the cache from the database. Notably, the generation of bloom filters for DQL queries can be seamlessly integrated with the scan operators within the database since the tuples traversed by these operators are precisely the data that should be inserted into the bloom filters. Additionally, the neighbors required for signatures in DML statements can be retrieved from the index during index updates. This design simplifies the solution significantly.

\subsection{Overhead on the Database and Cache Sides}
As discussed in Section~\ref{sec_perAnal}, the performance of an invalidation strategy depends on the DQL query cost $c_q'$, and the cost of DML statements $c_u'$. Notably, the overhead in queries primarily arises from the signature and index. To mitigate the impact of the index on the cache side, we have thoughtfully designed a Q-Tree for transparent-$\mathcal{P}$ and a BF-Tree for transparent-$\mathcal{B}$, respectively. Leveraging the same tree structure for both solutions, their overheads are expected to be competitive. Therefore, our discussion focuses on the overhead associated with the signature on the database side.

In the case of the transparent-$\mathcal{P}$ solution, the signatures of DQL queries are predicates, which are available from the query parser of databases and the overhead is negligible. However, the signature of DML statements may need to generate additional queries when handling join queries. Therefore, it is not recommended for use in scenarios characterized by explosive joins, where joins generate a substantial number of results for each updated tuple, consequently leading to a considerable overhead in signature generation and Q-tree searches.
In contrast, in the transparent-$\mathcal{B}$ solution, the overhead associated with signature generation is negligible due to the efficiency of bloom filters~\cite{BF}. 

\subsection{False Positive Rate}
In Section~\ref{sec_perAnal}, we anticipate that the false positive rate for transparent invalidation strategies should ideally be near 0.  
In the case of the transparent-$\mathcal{P}$ solution, its false positive rate is promising to 0  due to specific predicate judgment. However, due to the potential false positive judgments of bloom filters, achieving a false positive rate of 0 cannot be realistic for the  transparent-$\mathcal{B}$ solution. 

The bloom filter is well-known for its efficient use of space and time, albeit at the cost of potential false positives. The false positive rate $p$ can be calculated using the formula $p=(1-e^{-\frac{K\times N}{M}})^K$, where $M$ is its bit array size, $K$ is the number of hash functions, and $N$ denotes the number of elements inserted into it. When $M=128$ (i.e., 16B) and $N=5$, the false positive rate can be impressively low as less than $10^{-5}$ (with a proper $K$). However, as $N$ increases to 25, the false positive rate can surge to over 8\% (with all $K$s), which may be unacceptable in certain scenarios.

To address this situation, two potential solutions can be considered. Firstly, as the formula implies, maintaining a low false positive rate is achievable by increasing the bit array size $M$ in proportion to $N$. However, bloom filters with distinct configurations, resulting from different $M$ values, cannot be indexed using the same index since they set different bits to 1 for the same key. This necessitates the cache's ability to discern configurations for bloom filters and uphold a distinct index for each configuration. In addition, $K(U)$ must also work in tandem with these configurations to generate a bloom filter for each one. This would reintroduce a level of tight coupling between the cache and the database, which was an issue we sought to avoid. Hence, we further propose a second solution.

Typically, the database can generate multiple bloom filters for a single DQL query, effectively maintaining a low false positive rate. If $n$ bloom filters are generated for a query and the original false positive rate of each individual bloom filter is $p$, the overall false positive rate for the query is $1-(1-p)^n \approx n\times p$. For instance, when 5 bloom filters with $M=128$ are generated for a query that accesses 25 tuples, the overall false positive rate remains below $10^{-4}$, a level that is typically acceptable. It's worth noting that, to uphold a low false positive rate, there is an associated additional space cost for queries accessing a substantial number of tuples. However, we argue that such queries should be tolerant of a slightly higher false positive rate and allow for the inclusion of more tuples within a single bloom filter. Therefore, an upper bound for the space cost of bloom filters can be imposed.

\subsection{Summary}
In summary, the transparent-$\mathcal{P}$ solution promises superior performance but places constraints on the types of queries it can handle and requires tight cooperation between the database and the cache. In contrast, transparent-$\mathcal{B}$ offers a more general and decoupled solution, albeit with the potential risk of false invalidations.

\section{Evaluation} \label{Sec_eval}

According to the elaboration in Sections~\ref{Sec_Auto}, in theory, our proposed mechanism can achieve transparent cache invalidation. That is, once an update occurs to the backend database, it will automatically identify all the cache entries that should be invalidated. 
The remaining question is whether this mechanism is sufficiently cost-effective such that it can be applied to real-world applications. 

We conducted extensive experiments to evaluate the cost-effectiveness of our mechanism. 
We first provide an end-to-end evaluation to assess the effectiveness of the two specific transparent invalidation solutions at the application level. We show that, compared to manual invalidation approaches, they can improve the cache hit ratio and at the same time reduce the cases of stale reads. Following that, we discussed the overhead introduced by them on the database side and cache side respectively. We revealed that the overhead they introduced can be acceptable compared to their improvement in cache utilization.

\subsection{Setup and Workloads}

During our experiments, two machines were used. Machine A was operated by Ubuntu 20.04  LTS, equipped with 4 sockets, each containing 18 cores (Intel(R) Xeon(R) Gold 5318H CPU @ 2.50GHz), and 200GB DRAM. Machine B was also operated by Ubuntu 20.04  LTS, equipped with 2 sockets, each containing 16 cores (Intel(R) Xeon(R) Gold 6226R CPU @ 2.90GHz), and 100GB DRAM. For all experiments, only one socket in each machine was utilized, with CPU and memory resources being limited to the same socket.

To evaluate the overall effect of our scheme of transparent cache invalidation, we set up a framework like the one in Figure~\ref{framework}. We used PostgreSQL v12.16 as the database and Redis v6.2.6 as the cache. We leveraged extension functions of PostgreSQL to retrieve necessary information from the database. In the experiments, the cache and clients were deployed to machine A with the max available memory of Redis being set as 4GB. The database server ran in a docker on machine B with data stored on SSD and the available memory limited to 4GB.

In particular, we used the TPC-C~\cite{TPCC} and YCSB~\cite{YCSB} benchmarks to simulate the application workloads. For TPC-C, we used two types of transactions only, with 5\% new-order transactions to update the data and 95\% stock-level transactions to query the data. Besides, we tried two settings with the warehouse as 100 and 1000 respectively. For YCSB, we pre-loaded 10 million 1000-byte tuples into the database and tried three workloads with different request proportions as follows,  where the $\mathit{maxscanlength}$ parameter of range requests was 10:
\begin{itemize}
    \item YCSB-RH: 90\% point read, 5\% update, 5\% range scan.
    \item TCSB-SH: 50\% point read, 5\% update, 45\% range scan.
    \item YCSB-MIX: 50\% point read, 25\% update, 25\% range scan.
\end{itemize}

\subsection{Effects of Transparent Invalidation} \label{Sec_expOverhead}
Through automatic cache invalidation, our mechanism can relieve the burdens of software developers in cache management.
But how does it compare to traditional invalidation methods in cache efficiency?   
In our first set of experiments, we compared the effect of fine-grained cache invalidation, which is enabled by our auto-invalidation mechanism, against that of TTL and rule-based invalidation.
Specifically, we evaluated the following cache invalidation solutions:
\begin{itemize}
	\item \textbf{transparent-$\mathcal{P}$} represents our approach to precise and transparent cache invalidation with the predicates as signatures.
        \item \textbf{transparent-$\mathcal{B}$} represents our approach with the bloom filters as signatures.
	\item \textbf{coarse} represents the approach applying traditional table-level coarser-grained invalidation. 
	\item \textbf{TTL-$t$} represents the traditional approach that used TTL for cache invalidation, where the expiration time was  $t$.
\end{itemize}

%\subsubsection{strategies for comparison.}
%Our solution aims to alleviate the cache management burden for software developers. But how does it compare to traditional invalidation methods in cache efficiency? 

\begin{figure}[!t]
    \centering
    \includegraphics[width=\linewidth]{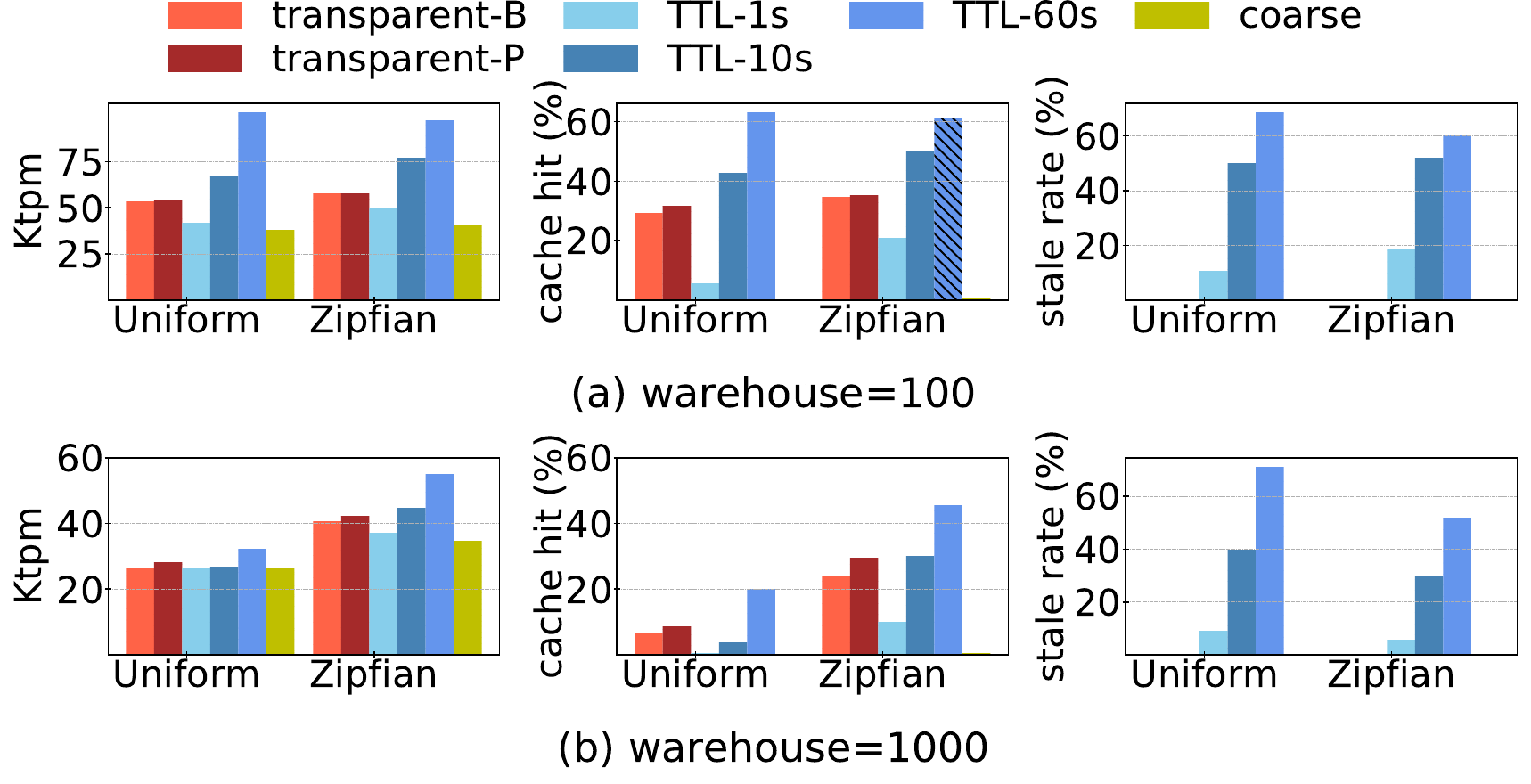}
    \caption{ Performance of different invalidation strategies on the TPC-C benchmark.}
    \label{tpcc}

\end{figure}

\begin{figure}[!t]
    \centering
    \includegraphics[width=\linewidth]{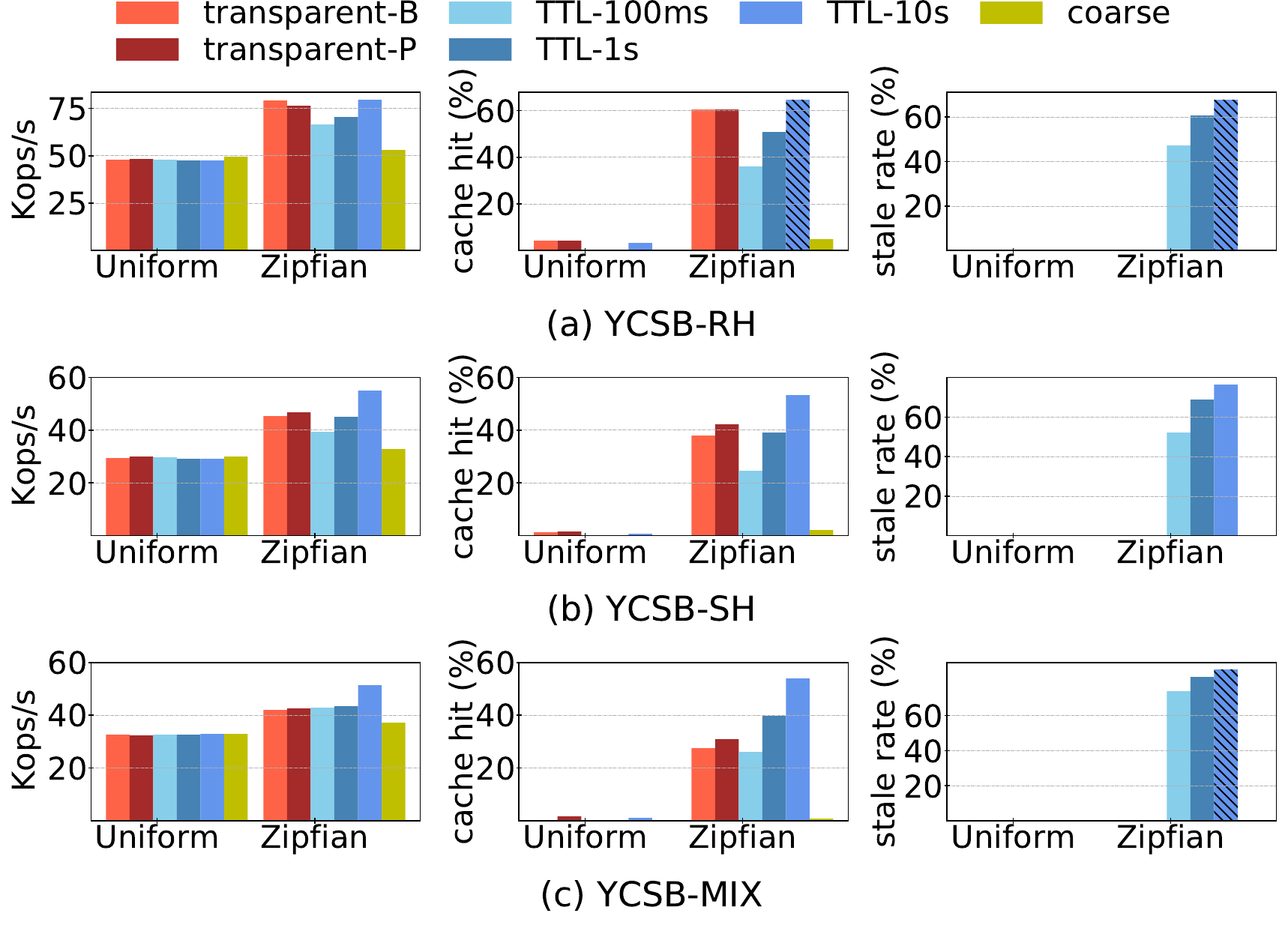}
    \caption{ Performance of different invalidation strategies on the YCSB benchmark.}
    \label{ycsb}

\end{figure}

In all the experiments, we employed 30 concurrent clients to carry out data accesses. We explored different data request distributions, including uniform and Zipfian (i.e., skewed) distributions. Specifically, the bit array size of bloom filters in the transparent-$\mathcal{B}$ strategy was set at 16 bytes for every 10 tuples in the YCSB benchmark and 512 bytes for every 200 tuples in the TPC-C benchmark. This setting was based on the different numbers of tuples accessed by the query.
Figure~\ref{tpcc} and Figure~\ref{ycsb} display the throughputs and cache hit ratios of different invalidation strategies. Notably, TTL-based strategies may result in stale reads. The figures also provide insights into the percentages of stale reads within cached hits.

The two benchmarks represent very different scenarios. The TPC-C benchmark contains more complex transactions, which entail relatively low cache hit rates. Nevertheless, as our transparent invalidation mechanism enables more precise cache management, as shown in Figure~\ref{tpcc}, it achieved better cache utility and, thus, superior performance compared to the coarse strategy and TTLs. Specifically, the coarse-grained approach was completely ineffective in this scenario. Its cache hit rates were too low to justify the adoption of application-level cache. A lengthier TTL, such as 10 seconds, could achieve a higher cache hit ratio but at the price of a higher stale rate (over 30\% in our experiments). As we discussed in Section~\ref{Sec_Survey}, controlling data freshness is always a difficult issue for applications applying the TTL strategy. As a result, the transparent invalidation strategies become more attractive. Besides, the results also show that the cache hit rate of transparent-$\mathcal{B}$ was a little lower than that of  transparent-$\mathcal{P}$. This is because of the false positive of bloom filters, which was about 0.8\% in our implementation with the 512-byte bloom filters for every 200 tuples.

In YCSB, the operations are simpler and faster. As a result, the TTL strategy can present a higher cache hit rate in some cases, but at the expense of high steal read rates, especially for the YCSB-MIX workloads.
As shown in Figure~\ref{ycsb}~(c), even when we decreased the TTL to as low as 1 second, the stale read rate remained over 60\% for Zipfian distributed. In particular, its stale rate can be low and negligible when data requests are uniformly distributed. However, the application-level cache seems to not work with the uniform distributed requests due to its extremely low cache hit rate. The figure also presents that, even in such scenarios, the transparent invalidation solutions invoked negligible overhead.

\subsection{Overhead Introduced to Database}
Transparent cache invalidation may introduce additional overheads to the database systems. This is particularly evident in the transparent-$\mathcal{P}$ solution, where the process of updating the database may lead to additional queries for generating signatures when dealing with join queries. To assess this impact, we further recorded the P95 latency for the new-order (i.e., update) and stock-level (i.e., read) transactions in the TPC-C benchmark and the update, point read, and range scan methods in the YCSB benchmark. 

\begin{figure}[!t]
    \centering
    \includegraphics[width=0.9\linewidth]{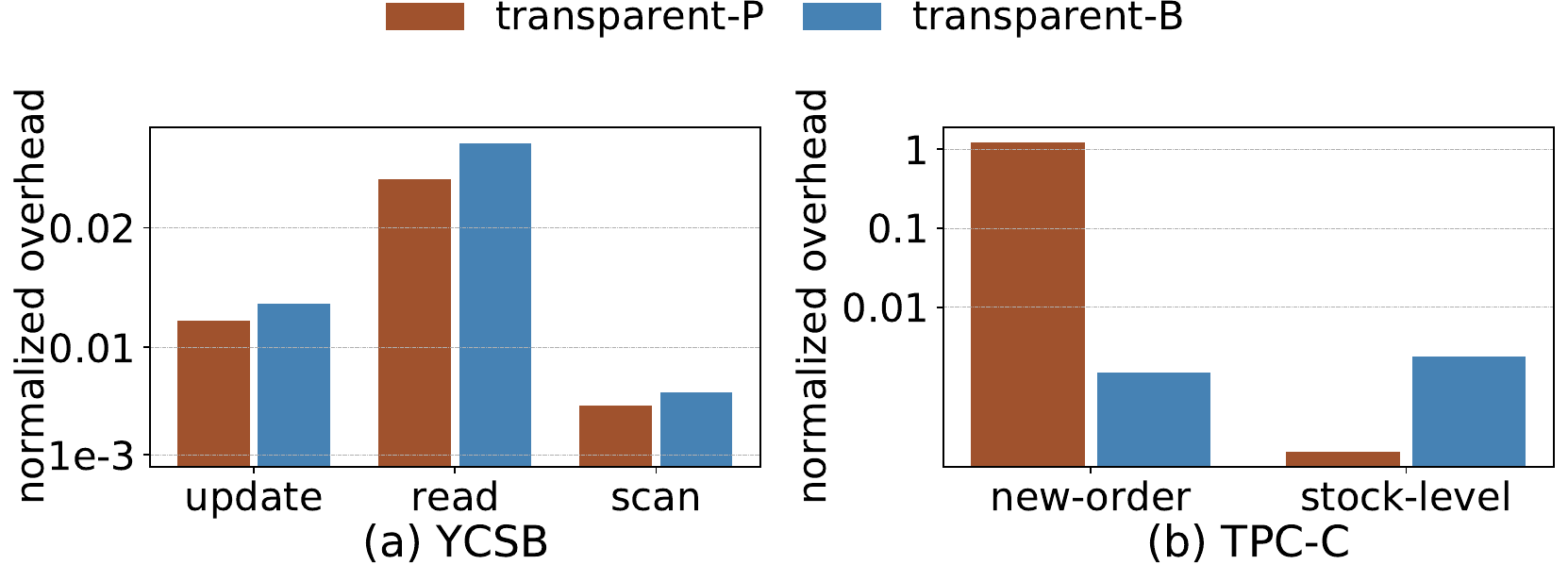}
    \caption{ Increase in P95 latency (in log scale) of database operations due to transparent invalidation strategies, normalized to that without transparent invalidation.}
    \label{overhead}

\end{figure}

Figure~\ref{overhead} presents the increases in P95 latency of different database operations due to the transparent invalidation strategies, which are normalized to the P95 latency of the original methods. 
Our results reveal that the influence of both solutions on end-to-end latency is minimal in the YCSB benchmark, amounting to less than 3\%. However, in the TPC-C benchmark, this impact is more pronounced for the transparent-$\mathcal{P}$ solution, with values of approximately 1.2x when the warehouse size is set at 1000. This discrepancy can be attributed to the presence of join queries in the TPC-C benchmark and the necessity of additional queries for signature generation during updates. Specifically, one update statement on the ``stock'' table in the new-order transaction can lead to over 10 joint tuples after invoking the join template. This leads to extremely high overhead on the signature generation. 

\subsection{Overhead Introduced to Cache} \label{Sec_expOpt}
%In this section, we mainly show the effects of using $removeLock$ and mark delete.
The index plays a central role in our cache system to achieve transparent cache invalidation. The cost-effectiveness of our approach depends on their overhead and efficiency. To evaluate the performance and overhead of our proposed indexes, we assessed them utilizing the YCSB benchmark carried out on machine A. 
In each experiment, we preloaded 10 million items into the index and issued concurrent {insertion} and {invalidation} requests to it. When an insertion causes the number of items in the index to reach the capacity limit (10 million), an {eviction} operation will be triggered. This simulates the cache replacement scenario.
We varied the proportion between insertion and invalidation to see the efficiency of the index with different workloads.
In particular, we evaluated the following indexes for comparison:
\begin{itemize}
        \item \textbf{Interval Tree} represents the classical interval tree based on a red-black tree~\cite{Algorithms}.
	\item \textbf{Q-Tree} represents the variant of interval tree introduced in Section~\ref{Sec_qtree}.
        \item \textbf{Trie Tree} represents the index for bloom filters as illustrated in Figure~\ref{fig:Trie}.
        \item \textbf{BF-Tree} represents our proposed index for bloom filters as illustrated in Figure~\ref{fig:BFTree}.
\end{itemize}

\begin{figure}[!t]
	\subfloat[Uniform distribution]{
		\begin{minipage}[t]{0.5\textwidth}
			\centering	
			\includegraphics[width=0.98\textwidth]{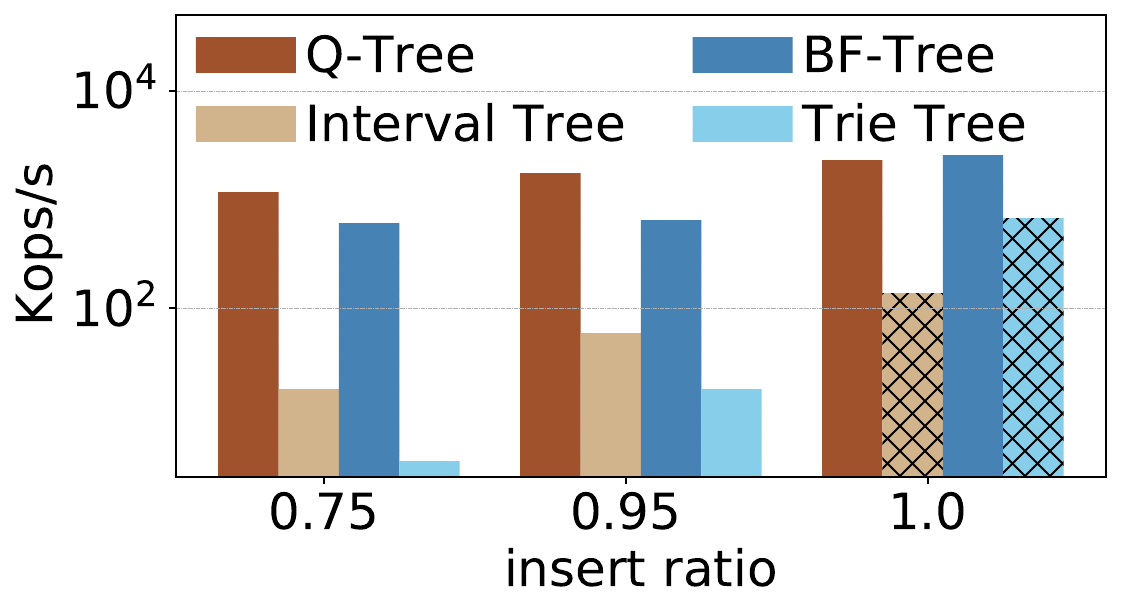}
		\end{minipage}
	}
	\subfloat[Zipfian distribution]{
		\begin{minipage}[t]{0.5\textwidth}
			\centering	
			\includegraphics[width=0.98\textwidth]{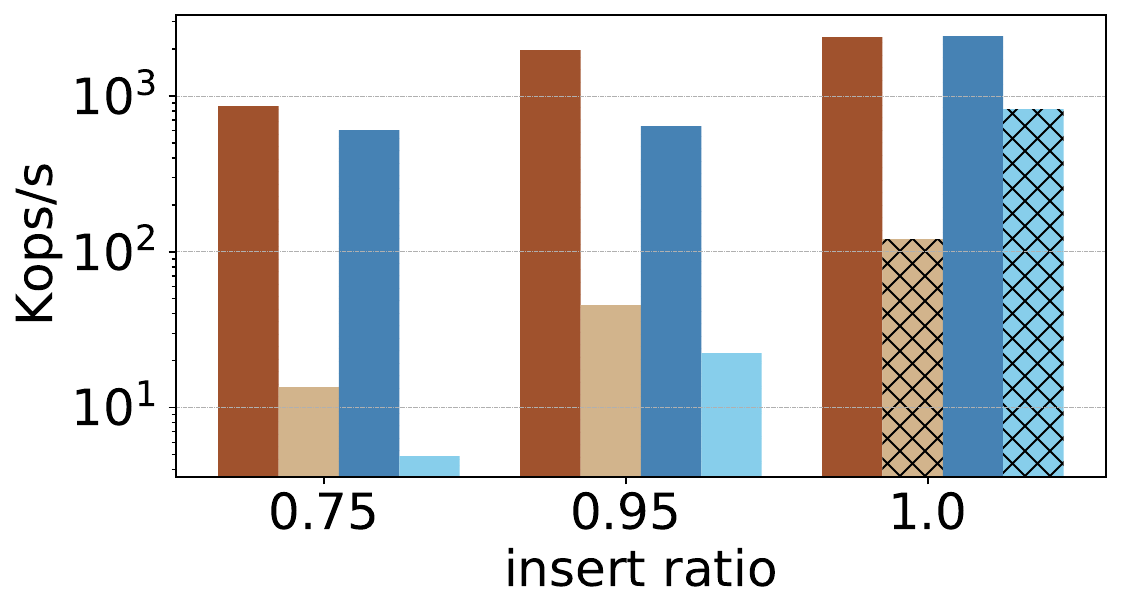}
		\end{minipage}
	} 
	\caption{Throughputs (in log scale) of different indexes with different insert ratios.}
	\label{time}	
\end{figure}

\begin{figure}[!t]
			\centering	
			\includegraphics[width=0.58\textwidth]{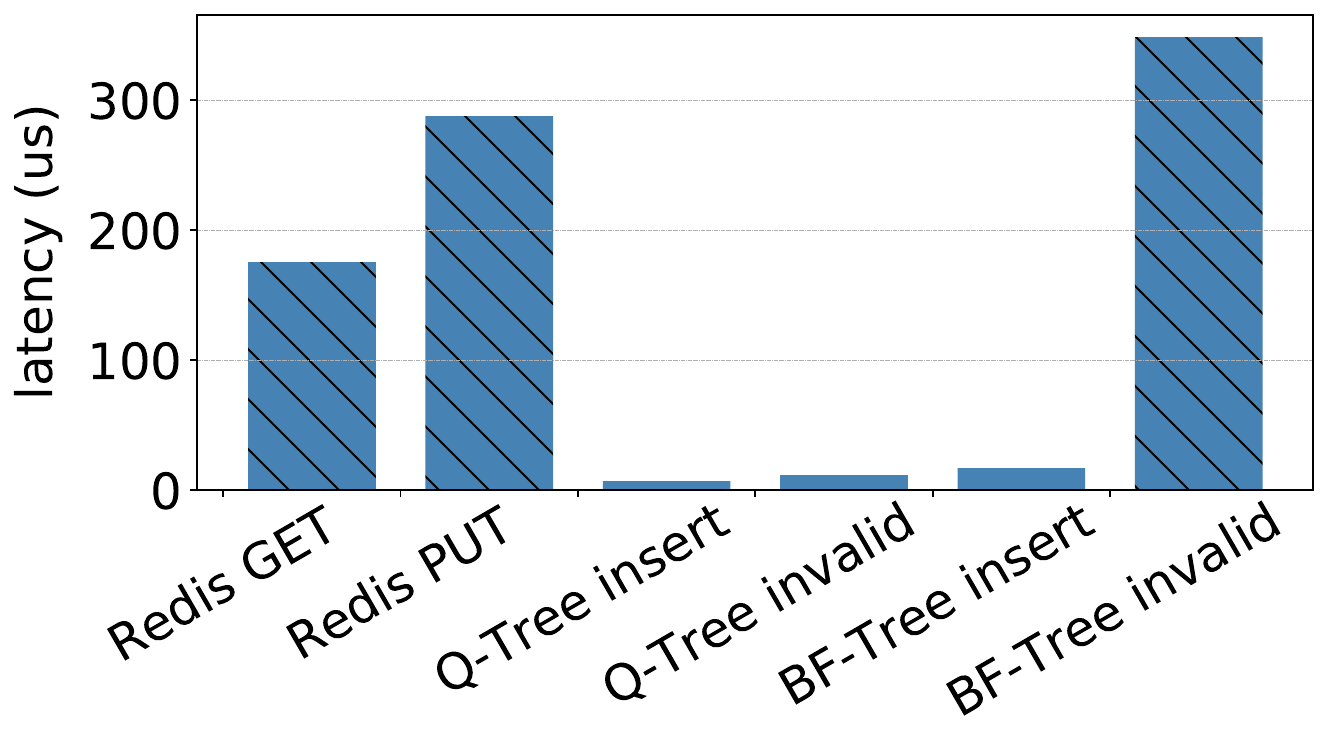}

	\caption{P95 Latency of different operations.}
	\label{late}	
\end{figure}

In our experiments, we utilized 16 concurrent client threads to issue requests, with the interval width ranging from 1 to 10. The bit array size in the BF-Tree was set to 16 bytes. Figure~\ref{time} illustrates the throughput of the four candidates with varying insertion-invalidation ratios and request distributions. (For example, if the insert ratio is 0.75, it indicates 75\% insertions and 25\% invalidations.)

As depicted in the figures, both the Q-Tree and BF-Tree significantly outperformed the baseline index, demonstrating their suitability for cache invalidation. Their overall performance proved to be more than adequate, with throughputs exceeding 500K OPS, and the Q-Tree even achieving over a million OPS, which surpasses the speed of both a typical database and a cache system. It's worth noting that, in our environment, the highest throughput of Redis was below 170K OPS, highlighting that the index will not become a bottleneck even under highly concurrent workloads.

Specifically, the memory footprints of both Q-Tree and BF-Tree were found to be less than 10\% of that required by Redis when we configured the cache entry size as 1000 bytes and the cache ID size as 20 bytes.
Additionally, Figure~\ref{late} provides a comparison of the P95 latency between Q-Tree and BF-Tree operations and the ``GET'' and ``PUT'' operations of Redis. As illustrated in the figure, the latency of Q-Tree operations is under 10\% of that of Redis operations. This highlights that the overhead incurred by Q-Tree is nearly negligible when compared to that of the cache.

The insertion operation of BF-Tree is competitive to that of Q-Trees, whereas its invalidation operation exhibits higher P95 latency. This is because of the lack of order in its keys different from that of Q-Tree. In BF-Tree, it's possible that the bitwise $or$ result of two children contains a given key, yet neither of them individually does. Consequently, BF-Tree may experience unnecessary node traversals.
It's important to note that the invalidation operation occurs as a result of database updates and is expected to be far less frequent than Redis operations. Therefore, the overhead associated with BF-Tree remains to be acceptable. Nonetheless, exploring specific optimizations for BF-Tree remains an intriguing and valuable topic for further study.

\subsection{Summary}

In conclusion, our experiments demonstrate that transparent cache invalidation can enhance cache utility, leading to overall performance improvements. While this approach introduces certain overheads, the automated and precise cache management offered by transparent cache invalidation can justify these costs. Particularly for applications prioritizing data freshness, transparent cache invalidation emerges as a highly recommended solution.

\section{Related Work} \label{sec_relat}

%\textbf{ Cache management middlewares.}
Cache management is a well-studied topic. While a number of tools have been invented for facilitating the management of application-level cache, they offer limited support for cache invalidation.
Some works explored approaches~\cite{AOP2006, qTemplate, Cachematic, Quaestor, SI_OSDI, UrWeb, MCC} to automatic cache invalidation or synchronization. They mainly rely on middleware to accomplish this task, which retrieves updated data items from the database and compares them against DQL queries to detect invalid cache entries. However, these approaches cannot deal with complex queries, such as those involving joins or range predicates. Few of them utilize indexes to speed up the invalidation. Instead of resorting to middleware, our framework of transparent cache invalidation decomposes the invalidation workflow into internal functions of the database and cache systems. Implementation inside the database exposes more information to the invalidation workflow, allowing it to deal with more complex queries. Moreover, we proposed an indexing scheme to facilitate invalidation, which works uniformly well on different query predicates.

%\textbf{ Query result cache \& materialized view.}
Various solutions \cite{Dbcache, MTCache, ChronoCache, Ghandeharizadeh19}  have been proposed to enable efficient and transparent maintenance of query result cache in database systems.
For instance, MySQL\footnote{\url{https://dev.mysql.com/doc/refman/5.6/en/query-cache.html}} and Oracle\footnote{\url{https://docs.oracle.com/database/121/TGDBA/tune\_result\_cache.htm\#TGDBA642}} offer built-in query result caches, while Mysqlnd\footnote{\url{https://dev.mysql.com/doc/connectors/en/apis-php-mysqlnd-qc.html}} and ProxySQL\footnote{\url{https://proxysql.com/}} are external query result caches to be deployed as middleware. 
The materialized view is also an alternative to query result cache, as it is also intended to support result reuse in data processing \cite{CachePortal, Couchbase}.
In particular, Noria \cite{Noria} is a novel system that applies materialized views rather than caches to enhance the performance of a web application.
However, both solutions are primarily designed to expedite database queries and may not directly apply to application-level cache, which operates outside the database system. %can save costs in both database systems and application servers.  

%\textbf{Techniques for automatic cache invalidation.} Cache invalidation has been an extensively studied topic~\cite{AOP2006, qTemplate, CacheGenie, Cachematic}. Only a limited number of works~\cite{SI_OSDI, UrWeb} considered using indexes to speed up invalidation. However, none of them take into full consideration of complex queries, including range, multicast, and join queries, which are the main challenges in automatic cache invalidation.

%\textbf{Data provenance.}
Data provenance (a.k.a. data lineage) \cite{dataProvenance, ProvenanceNext} is a series of techniques to trace data and its origins in complex transformations.
It has received attention in many fields, such as BlockChains \cite{provenanceblockchain} and IoTs \cite{provenanceIoT}.
As it is intended to link outputs with origins \cite{Auditing}, it can potentially be used for cache invalidation.
However, as traditional data provenance techniques are devised for completely different scenarios, we did not find any technique directly applicable to our case.
For instance, Titian \cite{Titian} is a Spark library to support data provenance.
It assigns a unique identifier to each pair of output and origin. While it can be used to trace the relationship between cache entries and database objects, its space cost is hardly acceptable for application-level caches.

\section{Outlook} \label{sec_conclu}
%This paper investigated the challenges of cache management by conducting qualitative studies on GitHub projects and an online survey of developers. To implement transparent application-level cache invalidation, we proposed a solution that enhances both the database and cache to automatically establish the links between the cache entries and update methods. A central element of this mechanism is a versatile index on the cache side. Our experimental results confirmed the effectiveness and efficiency of our solution.
In this paper, we addressed the prominent issue of cache invalidation in modern software development practices. To mitigate this challenge, we proposed a framework of transparent cache invalidation, necessitating modifying both the database and cache systems. We presented two specific solutions of the framework and provided the preliminary experimental results of them, which showcased the effectiveness of our proposed framework.

There are several promising directions for further research. Firstly, while our current solutions of transparent cache invalidation have shown potential, it is not without its limitations. For example, alternative approaches can be explored for indexing query signatures, as the efficiency of invalidation operations on Q-Tree and BF-Tree are important to the overhead of the framework and can be further optimized. %For example, if application issues only point queries to the database, a hash table could serve as a more cost-effective index than interval trees.

Secondly, optimization can be explored to minimize the overhead of signature generation on the database side, especially for the solution with predicates. As mentioned earlier, generating update signatures in this solution often involves executing a series of join queries or retrieving neighbors. We can potentially skip some of these queries if we know beforehand that their results will not trigger cache invalidation. Additionally, we can combine multiple queries to reduce execution costs.

Thirdly, we can also explore optimization opportunities on the cache side. As not all applications require data freshness to the same extent, striking the right tradeoff between transparent invalidation and TTL can be explored more deeply.



\begin{thebibliography}{36}
\ifx \bisbn   \undefined \def \bisbn  #1{ISBN #1}\fi
\ifx \binits  \undefined \def \binits#1{#1}\fi
\ifx \bauthor  \undefined \def \bauthor#1{#1}\fi
\ifx \batitle  \undefined \def \batitle#1{#1}\fi
\ifx \bjtitle  \undefined \def \bjtitle#1{#1}\fi
\ifx \bvolume  \undefined \def \bvolume#1{\textbf{#1}}\fi
\ifx \byear  \undefined \def \byear#1{#1}\fi
\ifx \bissue  \undefined \def \bissue#1{#1}\fi
\ifx \bfpage  \undefined \def \bfpage#1{#1}\fi
\ifx \blpage  \undefined \def \blpage #1{#1}\fi
\ifx \burl  \undefined \def \burl#1{\textsf{#1}}\fi
\ifx \doiurl  \undefined \def \doiurl#1{\url{https://doi.org/#1}}\fi
\ifx \betal  \undefined \def \betal{\textit{et al.}}\fi
\ifx \binstitute  \undefined \def \binstitute#1{#1}\fi
\ifx \binstitutionaled  \undefined \def \binstitutionaled#1{#1}\fi
\ifx \bctitle  \undefined \def \bctitle#1{#1}\fi
\ifx \beditor  \undefined \def \beditor#1{#1}\fi
\ifx \bpublisher  \undefined \def \bpublisher#1{#1}\fi
\ifx \bbtitle  \undefined \def \bbtitle#1{#1}\fi
\ifx \bedition  \undefined \def \bedition#1{#1}\fi
\ifx \bseriesno  \undefined \def \bseriesno#1{#1}\fi
\ifx \blocation  \undefined \def \blocation#1{#1}\fi
\ifx \bsertitle  \undefined \def \bsertitle#1{#1}\fi
\ifx \bsnm \undefined \def \bsnm#1{#1}\fi
\ifx \bsuffix \undefined \def \bsuffix#1{#1}\fi
\ifx \bparticle \undefined \def \bparticle#1{#1}\fi
\ifx \barticle \undefined \def \barticle#1{#1}\fi
\bibcommenthead
\ifx \bconfdate \undefined \def \bconfdate #1{#1}\fi
\ifx \botherref \undefined \def \botherref #1{#1}\fi
\ifx \url \undefined \def \url#1{\textsf{#1}}\fi
\ifx \bchapter \undefined \def \bchapter#1{#1}\fi
\ifx \bbook \undefined \def \bbook#1{#1}\fi
\ifx \bcomment \undefined \def \bcomment#1{#1}\fi
\ifx \oauthor \undefined \def \oauthor#1{#1}\fi
\ifx \citeauthoryear \undefined \def \citeauthoryear#1{#1}\fi
\ifx \endbibitem  \undefined \def \endbibitem {}\fi
\ifx \bconflocation  \undefined \def \bconflocation#1{#1}\fi
\ifx \arxivurl  \undefined \def \arxivurl#1{\textsf{#1}}\fi
\csname PreBibitemsHook\endcsname

%%% 1
\bibitem[\protect\citeauthoryear{Mertz and Nunes}{2018}]{Survey_Application}
\begin{barticle}
\bauthor{\bsnm{Mertz}, \binits{J.}},
\bauthor{\bsnm{Nunes}, \binits{I.}}:
\batitle{Understanding application-level caching in web applications: {A}
  comprehensive introduction and survey of state-of-the-art approaches}.
\bjtitle{{ACM} Comput. Surv.}
\bvolume{50}(\bissue{6}),
\bfpage{98}--\blpage{19834}
(\byear{2018})
\doiurl{10.1145/3145813}
\end{barticle}
\endbibitem

%%% 2
\bibitem[\protect\citeauthoryear{Mertz and Nunes}{2017}]{Study_Application}
\begin{barticle}
\bauthor{\bsnm{Mertz}, \binits{J.}},
\bauthor{\bsnm{Nunes}, \binits{I.}}:
\batitle{{A Qualitative Study of Application-Level Caching}}.
\bjtitle{TOSEM}
\bvolume{43}(\bissue{9}),
\bfpage{798}--\blpage{816}
(\byear{2017})
\end{barticle}
\endbibitem

%%% 3
\bibitem[\protect\citeauthoryear{Zakhary et~al.}{2017}]{VLDB_TUTORIAL}
\begin{barticle}
\bauthor{\bsnm{Zakhary}, \binits{V.}},
\bauthor{\bsnm{{El Abbadi}}, \binits{A.}},
\bauthor{\bsnm{Agrawal}, \binits{D.}}:
\batitle{Caching at the web scale}.
\bjtitle{{PVLDB}}
\bvolume{10}(\bissue{12}),
\bfpage{2002}--\blpage{2005}
(\byear{2017})
\doiurl{10.14778/3137765.3137831}
\end{barticle}
\endbibitem

%%% 4
\bibitem[\protect\citeauthoryear{Glasbergen et~al.}{2020}]{ChronoCache}
\begin{botherref}
\oauthor{\bsnm{Glasbergen}, \binits{B.}},
\oauthor{\bsnm{Langendoen}, \binits{K.}},
\oauthor{\bsnm{Abebe}, \binits{M.}},
\oauthor{\bsnm{Daudjee}, \binits{K.}}:
Chronocache: Predictive and adaptive mid-tier query result caching,
2391--2406
(2020)
\doiurl{10.1145/3318464.3380593}
\end{botherref}
\endbibitem

%%% 5
\bibitem[\protect\citeauthoryear{}{2023}]{Spring}
\begin{botherref}
Caching Data with Spring.
\url{https://spring.io/guides/gs/caching/}
(2023)
\end{botherref}
\endbibitem

%%% 6
\bibitem[\protect\citeauthoryear{}{2023}]{Django}
\begin{botherref}
Django's cache.
\url{https://docs.djangoproject.com/en/3.0/topics/cache/}
(2023)
\end{botherref}
\endbibitem

%%% 7
\bibitem[\protect\citeauthoryear{Mertz and Nunes}{2018}]{Mertz2018}
\begin{barticle}
\bauthor{\bsnm{Mertz}, \binits{J.}},
\bauthor{\bsnm{Nunes}, \binits{I.}}:
\batitle{{Automation of application-level caching in a seamless way}}.
\bjtitle{SPE}
\bvolume{48}(\bissue{6}),
\bfpage{1218}--\blpage{1237}
(\byear{2018})
\doiurl{10.1002/spe.2571}
\end{barticle}
\endbibitem

%%% 8
\bibitem[\protect\citeauthoryear{Gessert et~al.}{2017}]{Quaestor}
\begin{barticle}
\bauthor{\bsnm{Gessert}, \binits{F.}},
\bauthor{\bsnm{Schaarschmidt}, \binits{M.}},
\bauthor{\bsnm{Wingerath}, \binits{W.}},
\bauthor{\bsnm{Witt}, \binits{E.}},
\bauthor{\bsnm{Yoneki}, \binits{E.}},
\bauthor{\bsnm{Ritter}, \binits{N.}}:
\batitle{Quaestor: Query web caching for database-as-a-service providers}.
\bjtitle{{PVLDB}}
\bvolume{10}(\bissue{12}),
\bfpage{1670}--\blpage{1681}
(\byear{2017})
\doiurl{10.14778/3137765.3137773}
\end{barticle}
\endbibitem

%%% 9
\bibitem[\protect\citeauthoryear{Holmqvist et~al.}{2019}]{Cachematic}
\begin{botherref}
\oauthor{\bsnm{Holmqvist}, \binits{V.}},
\oauthor{\bsnm{Nilsfors}, \binits{J.}},
\oauthor{\bsnm{Leitner}, \binits{P.}}:
{Cachematic - Automatic Invalidation in Application-Level Caching Systems}.
ICPE,
167--178
(2019)
\doiurl{10.1145/3297663.3309666}
\end{botherref}
\endbibitem

%%% 10
\bibitem[\protect\citeauthoryear{Laigner
  et~al.}{2021}]{DBLP:journals/pvldb/LaignerZSLK21}
\begin{barticle}
\bauthor{\bsnm{Laigner}, \binits{R.N.}},
\bauthor{\bsnm{Zhou}, \binits{Y.}},
\bauthor{\bsnm{Salles}, \binits{M.A.V.}},
\bauthor{\bsnm{Liu}, \binits{Y.}},
\bauthor{\bsnm{Kalinowski}, \binits{M.}}:
\batitle{Data management in microservices: State of the practice, challenges,
  and research directions}.
\bjtitle{Proc. {VLDB} Endow.}
\bvolume{14}(\bissue{13}),
\bfpage{3348}--\blpage{3361}
(\byear{2021})
\doiurl{10.14778/3484224.3484232}
\end{barticle}
\endbibitem

%%% 11
\bibitem[\protect\citeauthoryear{Cormen et~al.}{2009}]{Algorithms}
\begin{bbook}
\bauthor{\bsnm{Cormen}, \binits{T.H.}},
\bauthor{\bsnm{Leiserson}, \binits{C.E.}},
\bauthor{\bsnm{Rivest}, \binits{R.L.}},
\bauthor{\bsnm{Stein}, \binits{C.}}:
\bbtitle{Introduction to Algorithms, Third Edition},
\bedition{3rd} edn.
\bpublisher{The {MIT} Press}, \blocation{???}
(\byear{2009})
\end{bbook}
\endbibitem

%%% 12
\bibitem[\protect\citeauthoryear{Fisher}{1986}]{Hilbert}
\begin{barticle}
\bauthor{\bsnm{Fisher}, \binits{A.J.}}:
\batitle{A new algorithm for generating hilbert curves}.
\bjtitle{Softw. Pract. Exp.}
\bvolume{16}(\bissue{1}),
\bfpage{5}--\blpage{12}
(\byear{1986})
\doiurl{10.1002/spe.4380160103}
\end{barticle}
\endbibitem

%%% 13
\bibitem[\protect\citeauthoryear{{Wu} and {Chang}}{2009}]{HilbertSplit}
\begin{barticle}
\bauthor{\bsnm{{Wu}}, \binits{C.-.}},
\bauthor{\bsnm{{Chang}}, \binits{Y.-.}}:
\batitle{Quad-splitting algorithm for a window query on a hilbert curve}.
\bjtitle{IET Image Processing}
\bvolume{3}(\bissue{5}),
\bfpage{299}--\blpage{311}
(\byear{2009})
\end{barticle}
\endbibitem

%%% 14
\bibitem[\protect\citeauthoryear{Bouchenak et~al.}{2006}]{AOP2006}
\begin{botherref}
\oauthor{\bsnm{Bouchenak}, \binits{S.}},
\oauthor{\bsnm{Cox}, \binits{A.L.}},
\oauthor{\bsnm{Dropsho}, \binits{S.G.}},
\oauthor{\bsnm{Mittal}, \binits{S.}},
\oauthor{\bsnm{Zwaenepoel}, \binits{W.}}:
Caching dynamic web content: Designing and analysing an aspect-oriented
  solution
\textbf{4290},
1--21
(2006)
\doiurl{10.1007/11925071\_1}
\end{botherref}
\endbibitem

%%% 15
\bibitem[\protect\citeauthoryear{Garrod et~al.}{2008}]{qTemplate}
\begin{barticle}
\bauthor{\bsnm{Garrod}, \binits{C.}},
\bauthor{\bsnm{Manjhi}, \binits{A.}},
\bauthor{\bsnm{Ailamaki}, \binits{A.}},
\bauthor{\bsnm{Maggs}, \binits{B.}},
\bauthor{\bsnm{Mowry}, \binits{T.}},
\bauthor{\bsnm{Olston}, \binits{C.}},
\bauthor{\bsnm{Tomasic}, \binits{A.}}:
\batitle{{Scalable query result caching for web applications}}.
\bjtitle{PVLDB}
\bvolume{1}(\bissue{1}),
\bfpage{550}--\blpage{561}
(\byear{2008})
\doiurl{10.14778/1453856.1453917}
\end{barticle}
\endbibitem

%%% 16
\bibitem[\protect\citeauthoryear{Glasbergen et~al.}{2018}]{Apollo}
\begin{botherref}
\oauthor{\bsnm{Glasbergen}, \binits{B.}},
\oauthor{\bsnm{Abebe}, \binits{M.}},
\oauthor{\bsnm{Daudjee}, \binits{K.}},
\oauthor{\bsnm{Foggo}, \binits{S.}},
\oauthor{\bsnm{Pacaci}, \binits{A.}}:
Apollo: Learning query correlations for predictive caching in geo-distributed
  systems,
253--264
(2018)
\doiurl{10.5441/002/edbt.2018.23}
\end{botherref}
\endbibitem

%%% 17
\bibitem[\protect\citeauthoryear{Lehman and Yao}{1981}]{blink}
\begin{barticle}
\bauthor{\bsnm{Lehman}, \binits{P.L.}},
\bauthor{\bsnm{Yao}, \binits{s.B.}}:
\batitle{Efficient locking for concurrent operations on b-trees}.
\bjtitle{ACM Trans. Database Syst.}
\bvolume{6}(\bissue{4}),
\bfpage{650}--\blpage{670}
(\byear{1981})
\doiurl{10.1145/319628.319663}
\end{barticle}
\endbibitem

%%% 18
\bibitem[\protect\citeauthoryear{Bloom}{1970}]{BF}
\begin{barticle}
\bauthor{\bsnm{Bloom}, \binits{B.H.}}:
\batitle{Space/time trade-offs in hash coding with allowable errors}.
\bjtitle{Commun. ACM}
\bvolume{13}(\bissue{7}),
\bfpage{422}--\blpage{426}
(\byear{1970})
\doiurl{10.1145/362686.362692}
\end{barticle}
\endbibitem

%%% 19
\bibitem[\protect\citeauthoryear{Black}{1998}]{algorithms2}
\begin{botherref}
\oauthor{\bsnm{Black}, \binits{P.E.}}:
Dictionary of algorithms and data structures
(1998)
\end{botherref}
\endbibitem

%%% 20
\bibitem[\protect\citeauthoryear{Scully and Chlipala}{2017}]{UrWeb}
\begin{barticle}
\bauthor{\bsnm{Scully}, \binits{Z.}},
\bauthor{\bsnm{Chlipala}, \binits{A.}}:
\batitle{{A program optimization for automatic database result caching}}.
\bjtitle{POPL}
\bvolume{52}(\bissue{1}),
\bfpage{271}--\blpage{284}
(\byear{2017})
\doiurl{10.1145/3093333.3009891}
\end{barticle}
\endbibitem

%%% 21
\bibitem[\protect\citeauthoryear{}{2023}]{TPCC}
\begin{botherref}
TPC-C.
\url{https://www.tpc.org/tpcc/}
(2023)
\end{botherref}
\endbibitem

%%% 22
\bibitem[\protect\citeauthoryear{Cooper et~al.}{}]{YCSB}
\begin{botherref}
\oauthor{\bsnm{Cooper}, \binits{B.F.}},
\oauthor{\bsnm{Silberstein}, \binits{A.}},
\oauthor{\bsnm{Tam}, \binits{E.}},
\oauthor{\bsnm{Ramakrishnan}, \binits{R.}},
\oauthor{\bsnm{Sears}, \binits{R.}}:
Benchmarking cloud serving systems with ycsb.
SoCC '10,
pp. 143--154.
\doiurl{10.1145/1807128.1807152} .
\url{https://doi.org/10.1145/1807128.1807152}
\end{botherref}
\endbibitem

%%% 23
\bibitem[\protect\citeauthoryear{Dan et~al.}{2010}]{SI_OSDI}
\begin{botherref}
\oauthor{\bsnm{Dan}, \binits{R.K.}},
\oauthor{\bsnm{Clements}, \binits{A.T.}},
\oauthor{\bsnm{Zhang}, \binits{I.}},
\oauthor{\bsnm{Madden}, \binits{S.}}:
{Transactional Consistency and Automatic Management in an Application Data
  Cache}.
OSDI
(2010)
\end{botherref}
\endbibitem

%%% 24
\bibitem[\protect\citeauthoryear{An and Cao}{2022}]{MCC}
\begin{barticle}
\bauthor{\bsnm{An}, \binits{S.}},
\bauthor{\bsnm{Cao}, \binits{Y.}}:
\batitle{Making cache monotonic and consistent}.
\bjtitle{Proc. {VLDB} Endow.}
\bvolume{16}(\bissue{4}),
\bfpage{891}--\blpage{904}
(\byear{2022})
\end{barticle}
\endbibitem

%%% 25
\bibitem[\protect\citeauthoryear{Altinel et~al.}{2002}]{Dbcache}
\begin{botherref}
\oauthor{\bsnm{Altinel}, \binits{M.}},
\oauthor{\bsnm{Luo}, \binits{Q.}},
\oauthor{\bsnm{Krishnamurthy}, \binits{S.}}:
{Dbcache: Database caching for web application servers}.
SIGMOD,
2001--2001
(2002)
\end{botherref}
\endbibitem

%%% 26
\bibitem[\protect\citeauthoryear{Larson et~al.}{2003}]{MTCache}
\begin{botherref}
\oauthor{\bsnm{Larson}, \binits{P.}},
\oauthor{\bsnm{Goldstein}, \binits{J.}},
\oauthor{\bsnm{Zhou}, \binits{J.}}:
Transparent mid-tier database caching in {SQL} server,
661
(2003)
\doiurl{10.1145/872757.872848}
\end{botherref}
\endbibitem

%%% 27
\bibitem[\protect\citeauthoryear{Ghandeharizadeh and
  Nguyen}{2019}]{Ghandeharizadeh19}
\begin{barticle}
\bauthor{\bsnm{Ghandeharizadeh}, \binits{S.}},
\bauthor{\bsnm{Nguyen}, \binits{H.}}:
\batitle{Design, implementation, and evaluation of write-back policy with cache
  augmented data stores}.
\bjtitle{Proc. {VLDB} Endow.}
\bvolume{12}(\bissue{8}),
\bfpage{836}--\blpage{849}
(\byear{2019})
\doiurl{10.14778/3324301.3324302}
\end{barticle}
\endbibitem

%%% 28
\bibitem[\protect\citeauthoryear{}{2001}]{CachePortal}
\begin{botherref}
{CachePortal Technology for Accelerating Database-driven E-commerce Web Sites}.
VLDB
(2001)
\end{botherref}
\endbibitem

%%% 29
\bibitem[\protect\citeauthoryear{Borkar et~al.}{2016}]{Couchbase}
\begin{botherref}
\oauthor{\bsnm{Borkar}, \binits{D.}},
\oauthor{\bsnm{Mayuram}, \binits{R.}},
\oauthor{\bsnm{Sangudi}, \binits{G.}},
\oauthor{\bsnm{Carey}, \binits{M.}}:
{Have Your Data and Query It Too}.
SIGMOD,
239--251
(2016)
\doiurl{10.1145/2882903.2904443}
\end{botherref}
\endbibitem

%%% 30
\bibitem[\protect\citeauthoryear{Gjengset et~al.}{2018}]{Noria}
\begin{botherref}
\oauthor{\bsnm{Gjengset}, \binits{J.}},
\oauthor{\bsnm{Schwarzkopf}, \binits{M.}},
\oauthor{\bsnm{Behrens}, \binits{J.}},
\oauthor{\bsnm{Timb}, \binits{L.}},
\oauthor{\bsnm{Ek}, \binits{M.}},
\oauthor{\bsnm{Kohler}, \binits{E.}},
\oauthor{\bsnm{Kaashoek}, \binits{M.F.}},
\oauthor{\bsnm{Morris}, \binits{R.}}:
{Noria: dynamic, partially-stateful data-flow for high-performance web
  applications}.
OSDI
(2018)
\end{botherref}
\endbibitem

%%% 31
\bibitem[\protect\citeauthoryear{Simmhan et~al.}{2005}]{dataProvenance}
\begin{barticle}
\bauthor{\bsnm{Simmhan}, \binits{Y.L.}},
\bauthor{\bsnm{Plale}, \binits{B.}},
\bauthor{\bsnm{Gannon}, \binits{D.}}, \betal:
\batitle{A survey of data provenance techniques}.
\bjtitle{Computer Science Department, Indiana University, Bloomington IN}
\bvolume{47405},
\bfpage{69}
(\byear{2005})
\end{barticle}
\endbibitem

%%% 32
\bibitem[\protect\citeauthoryear{Buneman and Tan}{2018}]{ProvenanceNext}
\begin{barticle}
\bauthor{\bsnm{Buneman}, \binits{P.}},
\bauthor{\bsnm{Tan}, \binits{W.}}:
\batitle{Data provenance: What next?}
\bjtitle{{SIGMOD} Rec.}
\bvolume{47}(\bissue{3}),
\bfpage{5}--\blpage{16}
(\byear{2018})
\doiurl{10.1145/3316416.3316418}
\end{barticle}
\endbibitem

%%% 33
\bibitem[\protect\citeauthoryear{Ruan et~al.}{2019}]{provenanceblockchain}
\begin{barticle}
\bauthor{\bsnm{Ruan}, \binits{P.}},
\bauthor{\bsnm{Chen}, \binits{G.}},
\bauthor{\bsnm{Dinh}, \binits{T.T.A.}},
\bauthor{\bsnm{Lin}, \binits{Q.}},
\bauthor{\bsnm{Ooi}, \binits{B.C.}},
\bauthor{\bsnm{Zhang}, \binits{M.}}:
\batitle{Fine-grained, secure and efficient data provenance on blockchain
  systems}.
\bjtitle{Proceedings of the VLDB Endowment}
\bvolume{12}(\bissue{9}),
\bfpage{975}--\blpage{988}
(\byear{2019})
\end{barticle}
\endbibitem

%%% 34
\bibitem[\protect\citeauthoryear{Kamal et~al.}{2018}]{provenanceIoT}
\begin{barticle}
\bauthor{\bsnm{Kamal}, \binits{M.}}, \betal:
\batitle{Light-weight security and data provenance for multi-hop internet of
  things}.
\bjtitle{IEEE Access}
\bvolume{6},
\bfpage{34439}--\blpage{34448}
(\byear{2018})
\end{barticle}
\endbibitem

%%% 35
\bibitem[\protect\citeauthoryear{Song and Shmatikov}{2019}]{Auditing}
\begin{botherref}
\oauthor{\bsnm{Song}, \binits{C.}},
\oauthor{\bsnm{Shmatikov}, \binits{V.}}:
Auditing data provenance in text-generation models,
196--206
(2019)
\doiurl{10.1145/3292500.3330885}
\end{botherref}
\endbibitem

%%% 36
\bibitem[\protect\citeauthoryear{Interlandi et~al.}{2018}]{Titian}
\begin{barticle}
\bauthor{\bsnm{Interlandi}, \binits{M.}},
\bauthor{\bsnm{Ekmekji}, \binits{A.}},
\bauthor{\bsnm{Shah}, \binits{K.}},
\bauthor{\bsnm{Gulzar}, \binits{M.A.}},
\bauthor{\bsnm{Tetali}, \binits{S.D.}},
\bauthor{\bsnm{Kim}, \binits{M.}},
\bauthor{\bsnm{Millstein}, \binits{T.D.}},
\bauthor{\bsnm{Condie}, \binits{T.}}:
\batitle{Adding data provenance support to apache spark}.
\bjtitle{{VLDB} J.}
\bvolume{27}(\bissue{5}),
\bfpage{595}--\blpage{615}
(\byear{2018})
\doiurl{10.1007/s00778-017-0474-5}
\end{barticle}
\endbibitem

\end{thebibliography}
\end{document}